\documentclass[12pt]{article}
\usepackage[utf8]{inputenc}
\usepackage[english]{babel}
\usepackage{a4wide}
\usepackage{amsfonts}
\usepackage{amssymb}
\usepackage{graphics,amsmath}
\usepackage{graphicx}
\usepackage{cancel}
\usepackage{cite}
\usepackage{mathrsfs}
\usepackage[usenames,dvipsnames,svgnames,table]{xcolor}
\usepackage{slashed}
\usepackage{etoolbox}
\usepackage{physics}

\def\hybrid{\topmargin -20pt    \oddsidemargin 0pt
        \headheight 0pt \headsep 0pt
        \textwidth 6.35in       % BS paper
        \textheight 9.25in       % BS paper
        \marginparwidth .875in
        \parskip 5pt plus 1pt   \jot = 1.5ex}

\hybrid

\def\baselinestretch{1.2}

\catcode`\@=11

\def\marginnote#1{}
\newcount\hour
\newcount\minute
\newtoks\amorpm
\hour=\time\divide\hour by60
\minute=\time{\multiply\hour by60 \global\advance\minute by-\hour}
\edef\standardtime{{\ifnum\hour<12 \global\amorpm={am}%
        \else\global\amorpm={pm}\advance\hour by-12 \fi
        \ifnum\hour=0 \hour=12 \fi
        \number\hour:\ifnum\minute<10 0\fi\number\minute\the\amorpm}}
\edef\militarytime{\number\hour:\ifnum\minute<10 0\fi\number\minute}
\def\draftlabel#1{{\@bsphack\if@filesw {\let\thepage\relax
   \xdef\@gtempa{\write\@auxout{\string
      \newlabel{#1}{{\@currentlabel}{\thepage}}}}}\@gtempa
   \if@nobreak \ifvmode\nobreak\fi\fi\fi\@esphack}
        \gdef\@eqnlabel{#1}}
\def\@eqnlabel{}
\def\@vacuum{}
\def\draftmarginnote#1{\marginpar{\raggedright\scriptsize\tt#1}}

\def\draft{\oddsidemargin -.5truein
        \def\@oddfoot{\sl preliminary draft \hfil
        \rm\thepage\hfil\sl\today\quad\militarytime}
        \let\@evenfoot\@oddfoot \overfullrule 3pt
        \let\label=\draftlabel
        \let\marginnote=\draftmarginnote
   \def\@eqnnum{(\theequation)\rlap{\kern\marginparsep\tt\@eqnlabel}%
\global\let\@eqnlabel\@vacuum}  }

\def\preprint{\twocolumn\sloppy\flushbottom\parindent 2em
        \leftmargini 2em\leftmarginv .5em\leftmarginvi .5em
        \oddsidemargin -.5in    \evensidemargin -.5in
        \columnsep .4in \footheight 0pt
        \textwidth 10.in        \topmargin  -.4in
        \headheight 12pt \topskip .4in
        \textheight 6.9in \footskip 0pt
        \def\@oddhead{\thepage\hfil\addtocounter{page}{1}\thepage}
        \let\@evenhead\@oddhead \def\@oddfoot{} \def\@evenfoot{} }

\def\numberbysection{\@addtoreset{equation}{section}
        \def\theequation{\thesection.\arabic{equation}}}

\def\underline#1{\relax\ifmmode\@@underline#1\else
        $\@@underline{\hbox{#1}}$\relax\fi}

\def\titlepage{\@restonecolfalse\if@twocolumn\@restonecoltrue\onecolumn
     \else \newpage \fi \thispagestyle{empty}\c@page\z@
        \def\thefootnote{\fnsymbol{footnote}} }

\def\endtitlepage{\if@restonecol\twocolumn \else \newpage \fi
        \def\thefootnote{\arabic{footnote}}
        \setcounter{footnote}{0}}  %\c@footnote\z@ }

\catcode`@=12
\relax

\def\figcap{\section*{Figure Captions\markboth
        {FIGURECAPTIONS}{FIGURECAPTIONS}}\list
        {Figure \arabic{enumi}:\hfill}{\settowidth\labelwidth{Figure
999:}
        \leftmargin\labelwidth
        \advance\leftmargin\labelsep\usecounter{enumi}}}
 \relax
\def\tablecap{\section*{Table Captions\markboth
        {TABLECAPTIONS}{TABLECAPTIONS}}\list
        {Table \arabic{enumi}:\hfill}{\settowidth\labelwidth{Table
999:}
        \leftmargin\labelwidth
        \advance\leftmargin\labelsep\usecounter{enumi}}}
 \relax
\def\reflist{\section*{References\markboth
        {REFLIST}{REFLIST}}\list
        {[\arabic{enumi}]\hfill}{\settowidth\labelwidth{[999]}
        \leftmargin\labelwidth
        \advance\leftmargin\labelsep\usecounter{enumi}}}
 \relax
\makeatletter
\newcounter{pubctr}
\def\publist{\@ifnextchar[{\@publist}{\@@publist}}
\def\@publist[#1]{\list
        {[\arabic{pubctr}]\hfill}{\settowidth\labelwidth{[999]}
        \leftmargin\labelwidth
        \advance\leftmargin\labelsep
        \@nmbrlisttrue\def\@listctr{pubctr}
        \setcounter{pubctr}{#1}\addtocounter{pubctr}{-1}}}
\def\@@publist{\list
        {[\arabic{pubctr}]\hfill}{\settowidth\labelwidth{[999]}
        \leftmargin\labelwidth
        \advance\leftmargin\labelsep
        \@nmbrlisttrue\def\@listctr{pubctr}}}
 \relax
\makeatother
\newskip\humongous \humongous=0pt plus 1000pt minus 1000pt

\newif\ifdtup

\relax

\def\be{\begin{equation}}
\def\ee{\end{equation}}
\def\ba{\begin{eqnarray}}
\def\ea{\end{eqnarray}}

\def\no{\noindent}

\def\IR{\relax{\rm I\kern-.18em R}}
\def\II{\relax{\rm 1\kern-.35em1}}

\renewcommand{\theequation}{\thesection.\arabic{equation}}
\csname @addtoreset\endcsname{equation}{section}

\def\IR{\relax{\rm I\kern-.18em R}}
\def\inv{^{\raise.15ex\hbox{${\scriptscriptstyle -}$}\kern-.05em 1}}

\DeclareMathOperator{\e}{e}

\DeclareMathOperator{\Background}{AdS_{3}\times S^{3}\times T^{4}}
\DeclareMathOperator{\Paradigm}{AdS_{5}\times S^{5}}
\DeclareMathOperator{\Manifold}{M_{3}}
\DeclareMathOperator{\dif}{d\!}

\DeclareMathOperator{\sn}{sn}

\DeclareMathOperator{\AdS}{AdS_{3}}
\DeclareMathOperator{\arccotan}{arccotan}

\DeclareMathOperator{\cn}{cn}
\DeclareMathOperator{\with}{with}

\allowdisplaybreaks

\begin{document}

%%%%%%%%%%%%%%%%%%%%%%%%%%%%%%%%%%%%%%%%%%%%%%%%%%%

\begin{titlepage}
\begin{center}

\vskip .5in

{\LARGE Double Yang-Baxter deformation of spinning strings} \\

\vskip 0.4in

{\bf Rafael Hern\'andez} \phantom{x} and \phantom{x} {\bf Roberto Ruiz} 
\vskip 0.1in

Departamento de F\'{\i}sica Te\'orica \\ and Instituto de F\'{\i}sica de Part\'{\i}culas y del Cosmos, IPARCOS \\
Universidad Complutense de Madrid \\
$28040$ Madrid, Spain \\

\vskip .2in

{\footnotesize{\tt rafael.hernandez@fis.ucm.es, roruiz@ucm.es}}

\end{center}

\vskip .4in

\centerline{\bf Abstract}
\vskip .1in
\no
\noindent
We study the reduction of classical strings rotating in the deformed three-sphere truncation of the double Yang-Baxter deformation of the $\Background$ background to an integrable mechanical model. 
The use of the generalized spinning-string ansatz leads to an integrable deformation of the Neumann-Rosochatius system. 
Integrability of this system follows from the fact that the usual constraints for the Uhlenbeck constants apply to any deformation that respects the isometric coordinates of the three-sphere. 
We construct solutions to the system in terms of the underlying ellipsoidal coordinate. 
The solutions depend on the domain of the deformation parameters and the reality conditions of the roots of a fourth order polynomial. 
We obtain constant-radii, giant-magnon and trigonometric solutions when the roots degenerate, and analyze the possible solutions in the undeformed limit.
In the case where the deformation parameters are purely imaginary and the polynomial involves two complex-conjugated roots, we find a new class of solutions. The new class is connected with twofold giant-magnon solutions in the degenerate limit of infinite period.

\vskip .4in
\noindent

\end{titlepage}

\vfill
\eject

\def\baselinestretch{1.2}

%%%%%%%%%%%%%%%%%%%%%%%%%%%%%%%%%%%%%%%%%%%%%%%%%%%%%%%%%%%%%%%%%%%%%%%%

\baselineskip 20pt

%%%%%%%%%%%%%%%%%%%%%%%%%%%%%%%%%%%%%%%%%%%%%%%%%%%%%%%%%%%%%%%%%%%%%%%%
%%%%%%%%%%%%%%%%%%%%%%%%%%%%%%%%%%%%%%%%%%%%%%%%%%%%%%%%%%%%%%%%%%%%%%%%

\section{Introduction}

The representation of classical strings rotating in the $\Paradigm$ background in terms of effective integrable mechanical systems is a renowned trait of the integrable structure underlying 
the AdS$_{5}$/CFT$_{4}$ correspondence~\cite{NR,NR2}. In reference~\cite{NR}, it was shown that the usage of a periodic ansatz for which the coordinates of the Cartan subalgebra of the $\Paradigm$ background 
are proportional to the world-sheet time reduces the action of the corresponding spinning string to the Neumann integrable system. The latter is a mechanical model that consists of a collection 
of $N+1$ simple harmonic oscillators restricted to lie in a $N$-dimensional sphere. The existence of the Uhlenbeck constants, a set of $N$ independent first integrals in involution, 
proves the integrability of the system by virtue of the Liouville theorem. The analysis was broadened in~\cite{NR2} by allowing non-trivial winding numbers along the compact coordinates of the Cartan subalgebra in the ansatz. 
The associated mechanical model is accordingly extended to the Neumann-Rosochatius integrable system, which involves an additional centrifugal potential for each oscillator. The integrability of the problem 
is preserved since the Uhlenbeck constants are also enhanced with terms that account for the centrifugal potentials. The connection between spinning strings and mechanical systems was extensively exploited 
in the study of the AdS$_{5}$/CFT$_{4}$ correspondence, as it permits to construct systematically classical solutions and their associated conserved charges by means of hyperelliptic functions and hyperelliptic integrals, respectively.

This picture poses the problem of the realization of such a connection in integrable deformations of backgrounds germane to the AdS$_{d+1}$/CFT$_{d}$ correspondence. The double Yang-Baxter deformation 
of the $\Background$ space-time supported with pure Ramond-Ramond three-form flux plays a distinctive role among this class~\cite{Lunin,Hoare,Seibold}. 
The permutation supercoset structure of the initial background allows an independent Yang-Baxter deformations for each of the two factors of the global symmetry group~\cite{Klimcik,biKlimcik}.
This type of two-parameter deformations were first considered in the context of the AdS$_{d+1}$/CFT$_{d}$ correspondence in~\cite{HRT}. There, the two-parameter deformation of the $\textnormal{O}\left(4\right)$ non-linear sigma-model put forward in~\cite{Fateev} was identified
with the double Yang-Baxter deformation of the $\textnormal{SU}\left(2\right)\times\textnormal{SU}\left(2\right)$ model of~\cite{Klimcik,biKlimcik}, 
and was then extended to the $\textnormal{AdS}_{3}\times\textnormal{S}^3$ space. Such a symmetric space was promoted to a full type IIB~superstring background in~\cite{Lunin} by means of the straightforward 
resolution of the supergravity equations. In reference~\cite{Hoare} an alternative background was obtained through a double Yang-Baxter deformation of the associated non-linear sigma 
model. Unlike the approach of \cite{Lunin}, the construction of~\cite{Hoare} ensures that the classical integrable structure is preserved. The invariance under kappa-symmetry transformations of the 
action was proven in~\cite{Seibold}, where the expressions for the Ramond-Ramond fluxes were also provided. From the viewpoint of the background fields, the constructions of both~\cite{Lunin} and~\cite{Hoare,Seibold} provide the same metric and dilaton, and an ignorable Kalb-Ramond two-form, 
but their respective Ramond-Ramond fluxes differ.~\footnote{Further developements in the context of double Yang-Baxter deformations and generalizations thereof may be found in \cite{Delduc1,Delduc2,Klimcik1,Klimcik2,Delduc3,Araujo,Three}. 
Moreover, classical solutions in this scenario have been considered in \cite{WK}.}

The aim of the present article is to analyze the extension of the Neumann-Rosochatius integrable system for spinning strings rotating in the deformed three-sphere truncation of the double Yang-Baxter deformation of the $\Background$ background. 
The body of the text is organized as follows. In section~\ref{Dos} we will put forward the deformation of the integrable mechanical model. 
We will prove the integrability of the model adducing the conservation of the canonically conjugate momenta to the angular coordinates and the Hamiltonian. We will then derive the deformation 
of the Uhlenbeck constants under the assumption that their initial constraints are respected by the deformation. The argument applies to every deformation of the three-sphere that upholds the isometric directions of the Cartan subalgebra. 
In section~\ref{Tres} we will construct formally the general solution to the deformation of the Neumann-Rosochatius system in terms of the ellipsoidal coordinate that parameterizes the equations of motion. 
The squared derivative of the ellipsoidal coordinate equals a fourth-order polynomial, whose degree is increased by one due to the presence of the deformation. We will show that the solutions depend 
on the domain of the parameters and the hierarchy of roots, obtaining the degenerate and undeformed limits in each case. 
We will find that the degenerate limits of the solution are realized by constant-radii and giant-magnon solutions, some of which have not undeformed counterpart.
In the case of purely imaginary deformation parameters, we will further
obtain trigonometric degenerate limits, and we will
construct a new class of 
general rotating
solutions. 
We will argue that this class reduce to twofold giant-magnon solutions in the degenerate limit of infinite period.
In section~\ref{Cuatro} we will conclude with a summary and comment on possible lines for further research.

%%%%%%%%%%%%%%%%%%%%%%%%%%%%%%%%%%%%%%%%%%%%%%%%%%%%%%%%%%%%%%%%%%%%%%%%
%%%%%%%%%%%%%%%%%%%%%%%%%%%%%%%%%%%%%%%%%%%%%%%%%%%%%%%%%%%%%%%%%%%%%%%%

\section{Deformation of the Neumann-Rosochatius system}

\label{Dos}

%This deformation was first proposed in terms of a two-parameter deformation of the $\textnormal{O}\left(4\right)$ non-linear sigma model in~\cite{Fateev}, and was later identified in~\cite{HRT} with a double Yang-Baxter deformation of the $\textnormal{SU}\left(2\right)\times\textnormal{SU}\left(2\right)$ non-linear sigma model~\cite{Klimcik,biKlimcik}. 

\noindent 
In this section we will present the extension of the Neumann-Rosochatius integrable system for spinning strings whose motion is confined 
to the two-parameter deformation of the three-sphere $\Manifold$.
We stress that the metric of $\Manifold$ can be obtained from the metric of the deformation of the $\Background$ space-time, since the double Yang-Baxter deformation does not mingle its $\textnormal{S}^{3}$-components with its $\AdS$-components or its $\textnormal{T}^4$-components~\cite{Lunin,Hoare,Seibold}. 
The metric of $\Manifold$ may be conveniently expressed in terms of the embedding coordinates of the three-sphere into $\mathbb{C}^{2}$. If the latter are denoted by $Z_{1}$ and $Z_{2}$, it reads~\cite{Hoare,Seibold} 
\be
\label{cirteM}
\begin{split}
\dif s^{2} & = \frac{1}{1+\kappa_{-}^{2}\bar{Z}_{1}Z_{1}+\kappa_{+}^{2}\bar{Z}_{2}Z_{2}} \Big(\dif \bar{Z}_{1}\dif Z_{1}+\dif \bar{Z}_{2}\dif Z_{2} \\
&- \frac{1}{4}\left[\kappa_{-}\left(Z_{1}\dif \bar{Z}_{1}- \bar{Z}_{1}\dif Z_{1}\right)
+\kappa_{+}\left(Z_{2}\dif \bar{Z}_{2}- \bar{Z}_{2}\dif Z_{2}\right)\right]^2\vphantom{\frac{1}{4}}\Big) \ ,
\end{split}
\ee
where $\kappa_{-}$ and $\kappa_{+}$ are the independent deformation parameters, the bar refers to complex conjugation and the embedding coordinates must be supplied with the constraint
\be
\label{tniartsnoC}
\bar{Z}_{1}Z_{1} + \bar{Z}_{2}Z_{2} = 1 \ .
\ee
The attributes of strict positivity and regularity of the metric demand either that the parameters belong to the real line or that they are purely imaginary numbers with modulus no greater 
than the unity, exclusive of the situation where both moduli equal one~\cite{Hoare}. The metric of the round three-sphere in complex embedding coordinates is retrieved when $\kappa_{\pm}=0$. 
The construction of the classical mechanical system of interest requires us to introduce the parameterization of the two complex coordinates
\be
Z_{a} = r_{a}\e^{i\varphi_{a}} \ , 
\ee 
with $a=1,2$. In these coordinates, the metric (\ref{cirteM}) turns into 
\be
\label{Metric}
\begin{split}
\dif s^{2}&=\frac{1}{1+\kappa_{-}^{2}r^{2}_{1}+\kappa_{+}^{2}r^{2}_{2}} 
\big[\dif r_{1}^2+\dif r_{2}^2+r_{1}^2\dif\varphi_{1}^2+r_{2}^2\dif\varphi_{2}^2+\left(\kappa_{-}r_{1}^2\dif\varphi_{1}+\kappa_{+}r_{2}^2\dif\varphi_{2}\right)^2\big] \ ,
\end{split}
\ee
and the constraint (\ref{tniartsnoC}) becomes
\be
\label{Constraint}
r_{1}^{2} + r_{2}^{2} = 1 \ .
\ee
In this parameterization, $\varphi_{a}$ are the azimuthal angles, with $0 \le \varphi_{a} < 2\pi$, 
whereas the radial coordinates $r_{a}$ are related to the polar angle of the round three-sphere via $r_{1} = \cos\theta$ and $r_{2} = \sin\theta$, with $0 \le \theta \le \pi$.

The starting point of our analysis is the Polyakov action for closed strings propagating in~$\mathbb{R}\times\Manifold$. The real line represents the time direction along the center of the deformed three-dimensional anti-de Sitter space 
and it is parameterized by the coordinate $t$. It remains underformed, since setting the hyperbolic radius in the global chart of the deformed anti-de Sitter space to zero erases the presence of the deformation parameters.
In the conformal gauge the action reads 
\be
\label{Action}
S = - \frac{\sqrt{\lambda}}{4\pi}\int_{-\infty}^{\infty}\dif\tau\int_{0}^{2\pi}\dif\sigma\left(-\eta^{\alpha\beta}\partial_{\alpha}t\partial_{\beta}t + 2L \right) \ ,
\ee
where $L$ is the non-linear sigma model Lagrangian corresponding to (\ref{Metric}). Note that, in principle, we should furnish the action above with both a Wess-Zumino term, 
which accounts for the Kalb-Ramond two-form, and a two-dimensional Hilbert-Einstein term, which couples the non-trivial dilaton of the background. 
In the case at issue, however, we may ignore the former, because the $B$-field is an exact two-form whose contribution vanishes by virtue of the periodic boundary conditions that closed strings exhibit. 
Neither the latter term is necessary, since the scalar of curvature of the world-sheet vanishes once we impose the conformal gauge.

In order to obtain the deformed Neumann-Rosochatius system we will impose the generalized spinning-string ansatz 
first proposed in~\cite{NR2}, which extends the Neumann system for spinning strings put forward in~\cite{NR}. The ansatz is
\be
t \left( \tau, \sigma\right ) = \kappa\tau \ , \quad r_{a}\left(\tau,\sigma\right) = r_{a}\left(\sigma\right) \ , \quad \varphi_{a} \left(\tau,\sigma\right) = \omega_{a}\tau+\alpha_{a}\left(\sigma\right) \ ,
\ee
where $\omega_{a}$ are the frequencies, and it is supplied with periodic boundary conditions pertinent to closed strings,
\be
\label{Parallel}
r_{a}\left(\sigma+2\pi\right)=r_{a}\left(\sigma\right) \ , \quad \alpha_{a}\left(\sigma+2\pi\right)=\alpha_{a}\left(\sigma\right)+2\pi m_{a} \ , \quad m_{a}\in\mathbb{Z} \ ,
\ee
where $m_{a}$ are the winding numbers. The usage of this ansatz reduces the Lagrangian in (\ref{Action}) to that of a classical mechanical system, 
\be
\label{Lagrangian}
\begin{split}
L & =\frac{1}{2\left(1+\kappa_{-}^{2}r^{2}_{1}+\kappa_{+}^{2}r^{2}_{2}\right)}\left[r'^{2}_{1}+r'^{2}_{2}+r_{1}^{2}\left(1+\kappa_{-}^2r_{1}^2\right) \left( - \omega_{1}^2+\alpha_{1}'^{2}  \right) + r_{2}^{2} \left(1+\kappa_{+}^2r_{2}^2\right)\right. \\
& \left.\times\left(- \omega_{2}^2+\alpha_{2}'^{2}  \right) + 2 \kappa_{-}\kappa_{+}r_{1}^2r_{2}^2\left(- \omega_{1}\omega_{2} + \alpha_{1}'\alpha_{2}' \right)\right]-\frac{\Lambda}{2}\left(r_{1}^2+r_{2}^2-1\right) \ ,
\end{split}
\ee
where the prime denotes derivative with respect to the spatial world-sheet coordinate, the constraint (\ref{Constraint}) has been implemented by means of the Lagrange multiplier $\Lambda$ and a superfluous prefactor has been omitted.  
The Virasoro constraints are accordingly written down as
\begin{align}
& r'^{2}_{1}+r'^{2}_{2}+r_{1}^{2}\left(1+\kappa_{-}^2r_{1}^2\right)\left(\omega_{1}^2+\alpha_{1}'^{2}\right)+r_{2}^{2}\left(1+\kappa_{+}^2r_{2}^2\right)\left(\omega_{2}^2+\alpha_{2}'^{2}\right) \nonumber \\
& + \, 2\kappa_{-}\kappa_{+}r_{1}^2r_{2}^2\left(\omega_{1}\omega_{2}+\alpha_{1}'\alpha_{2}'\right) =\left(1+\kappa_{-}^{2}r^{2}_{1}+\kappa_{+}^{2}r^{2}_{2}\right)\kappa^2 \ ,  \\
& r_{1}^{2}\left(1+\kappa_{-}^2r_{1}^2\right)\omega_{1}\alpha_{1}'+r_{2}^2\left(1+\kappa_{+}^2r_{2}^2\right)\omega_{2}\alpha_{2}'+\kappa_{+}\kappa_{-}r_{1}^2r_{2}^2\left(\omega_{1}\alpha_{2}'+\alpha_{1}'\omega_{2}\right)=0 \ . \nonumber
\end{align}
Besides, the invariance of the action under shifts along the directions of $t$ and $\varphi_{a}$ implies the conservation of the energy $E$ and the angular momenta~$J_{a}$, respectively. These conserved charges read
\be
\begin{split}
\label{Conserved}
E & =  -\sqrt{\lambda} \, \kappa \ , \\
J_{1} & = \frac{\sqrt{\lambda}}{2\pi}\int_{0}^{2\pi}\dif\sigma\frac{r_{1}^{2}\left(1+\kappa_{-}^2r_{1}^2\right)\omega_{1}+\kappa_{-}\kappa_{+}r_{1}^2r_{2}^2\omega_{2}}{1+\kappa_{-}^{2}r^{2}_{1}+\kappa_{+}^{2}r^{2}_{2}} \ , \\
J_{2} & =  \frac{\sqrt{\lambda}}{2\pi}\int_{0}^{2\pi}\dif\sigma\frac{r_{2}^{2}\left(1+\kappa_{+}^2r_{2}^2\right)\omega_{2}+\kappa_{-}\kappa_{+}r_{1}^2r_{2}^2\omega_{1}}{1+\kappa_{-}^{2}r^{2}_{1}+\kappa_{+}^{2}r^{2}_{2}} \ .
\end{split}
\ee

The cyclicity of the generalized coordinates $\alpha_{a}$ in the Lagrangian entails that their respective conjugate momenta $v_{a}$, explicitly expressed as
\be
\label{momenta}
v_{1}=\frac{r_{1}^2\left[\left(1+\kappa_{-}^2r_{1}^2\right)\alpha_{1}'+\kappa_{-}\kappa_{+}r_{2}^2\alpha_{2}'\right]}{1+\kappa_{-}^2r_{1}^2+\kappa_{+}^2r_{2}^2} \ , 
\quad v_{2}=\frac{r_{2}^2\left[\left(1+\kappa_{+}^2r_{2}^2\right)\alpha_{2}'+\kappa_{-}\kappa_{+}r_{1}^2\alpha_{1}'\right]}{1+\kappa_{-}^2r_{1}^2+\kappa_{+}^2r_{2}^2} \ ,
\ee
are conserved. The inversion of the previous relations leads to
\be
\alpha_{1}'=\frac{\left(1+\kappa_{+}^2r_{2}^2\right)v_{1}-\kappa_{-}\kappa_{+}r_{1}^2 v_{2}}{r_{1}^2} \ , \quad \alpha_{2}'=\frac{\left(1+\kappa_{-}^2r_{1}^2\right)v_{2}-\kappa_{-}\kappa_{+}r_{2}^2 v_{1}}{r_{2}^2} \ ,
\ee
whose integration provides an expression for the winding numbers $m_{a}$. We may then write down the Hamiltonian following from (\ref{Lagrangian}) like
\be
\label{Hamiltonian}
\begin{split}
H & = \frac{1}{2}\left[\frac{r_{1}'^2+r_{2}'^2}{1+\kappa_{-}^2r_{1}^2+\kappa_{+}^2r_{2}^2}+\omega_{1}^2r_{1}^2+\omega_{2}^2r_{2}^2 
- \frac{\left(\kappa_{+}\omega_{1}-\kappa_{-}\omega_{2}\right)^2r_{1}^2r_{2}^2}{1+\kappa_{-}^2r_{1}^2 + \kappa_{+}^{2}r_{2}^2} \right. \\
& \left. 
+ \frac{1+\kappa_{+}^2r_{2}^2}{r_{1}^2}v_{1}^2 + \frac{1+\kappa_{-}^2r_{2}^2}{r_{2}^{2}}v_{2}^2 - 2\kappa_{-}\kappa_{+} v_{1}v_{2} + \Lambda\left(r_{1}^2+r_{2}^2-1\right)\right] \ .
\end{split}
\ee
In the Hamiltonian picture, the Virasoro constraints are reformulated simply as
\be
\label{Virasoro}
2H=\kappa^2 \ , \quad \left(1+\kappa_{-}^2r_{1}^2+\kappa_{+}^2r_{2}^2\right)\left(\omega_{1}v_{1}+\omega_{2}v_{2}\right) = 0 \ .
\ee
While the first Virasoro constraint relates the energy $E$ with the angular momenta $J_{a}$ and the remaining parameters involved, the second constraint restricts the admissible values for these parameters. 
In view of the presence of three integrals of motion, $v_{a}$ and $H$, for the four generalized coordinates together with a constraint, we conclude that the two-parameter deformation of the Neumann-Rosochatius system 
preserves its classical integrability. This is of course expected since the initial non-linear sigma model has undergone two Yang-Baxter deformations, which uphold the classical integrable structure of the setting.

The integrable mechanical model above possesses some significant limits that are worth to underline. First and foremost, the undeformed Neumann-Rosochatius system~\cite{NR2} is recovered when we set $\kappa_{\pm}=0$. 
We can also retrieve a truncation of the $\eta$-deformation of the Neumann-Rosochatius system~\cite{NReta,AHM1,HNeta}, which describes classical strings spinning 
in the $\eta$-deformation of the $\Paradigm$ background, for $\kappa_{-}=0$ and $\kappa_{+}=\kappa$. Such a limit corresponds to the case in which the two factors of the global symmetry group 
of the semisymmetric space are deformed in the same way~\cite{Hoare,Seibold}. This implies that all considerations regarding the limits $\kappa=i$ and $\kappa=\infty$ studied in~\cite{HNeta} can be recovered 
starting from (\ref{Lagrangian}) for special choices of the deformation parameters. In particular, the 
limit $\kappa=i$ deserves a special mention, since in this case the spinning string rotates in a 
truncation of a ten-dimensional pp-wave background \cite{HRT} and it is possible to find explicit solutions to the 
Neumann-Rosochatius system in terms of hyperbolic functions~\cite{HNeta}.

\subsection{Integrals of motion}

The integrability of the $N$-dimensional undeformed Neumann system follows from the conservation of $N-1$ independent integrals of motion, known as the Uhlenbeck constants~\cite{NR}.~\footnote{Henceforth, the integer $N>1$ denotes the number of non-cyclic generalized coordinates. If $N=1$, the Uhlenbeck constants of the deformed Neumann-Rosochatius system may be postulated directly \cite{Roychowdhury}.}
The integrability of the $N$-dimensional Neumann-Rosochatius system is derived from the existence of $N$ first integrals analogue to $v_{a}$ for the cyclic coordinates and $N-1$ independent extended Uhlenbeck constants, 
which can be obtained from the Uhlenbeck integrals of the $2N$-dimensional Neumann system. Therefore, we may address the problem of proving the integrability of a deformation of the $N$-dimensional Neumann-Rosochatius system, 
provided that it preserves the directions of the Cartan subalgebra of the $(2N-1)$-dimensional sphere, attaining a deformation of the Uhlenbeck constants. This task may be fulfilled, for instance, trying to deform the Uhlenbeck constants 
directly~\cite{HNelliptic,HNeta} or drawing on the associated Lax connection in order to generate them~\cite{NReta,AHM1}. Nevertheless, the problem we study here is simplified by the low dimensionality of the model, 
which allows us to construct straightforwardly these first integrals in a systematic and symmetric fashion. 

Let us then consider the undeformed $N=2$ Neumann-Rosochatius system, which corresponds to setting $\kappa_{\pm}=0$ in the Hamiltonian (\ref{Hamiltonian}). 
If we solve the constraint (\ref{Constraint}), using for instance that $r_{1} = \cos \theta$ and $r_{2} =\sin \theta$, or, equivalently, restricting the coordinates of the phase space to lie upon the constraint submanifold, 
the Lagrangian multiplier is erased in (\ref{Hamiltonian}) and the Uhlenbeck constants satisfy
\be
\label{Redundancy}
H = \frac{1}{2}\left(\omega_{1}^2I_{1}+\omega_{2}^2I_{2}+v_{1}^2+v_{2}^2\right) \ , \quad I_{1}+I_{2}=1  \ .
\ee
Since $H$ is conserved, so they are the integrals $I_{1}$ and $I_{2}$. If the two squared frequencies $\omega_{a}^2$ differ between them (otherwise one would be dealing with a degenerate limit of the model connected with the Rosochatius model), 
one can express the two Uhlenbeck constants in terms of the Hamiltonian
\be
\label{Definition}
I_{1} = \frac{2H-\omega_{2}^2-v_{1}^2-v_{2}^2}{\omega_{1}^2-\omega_{2}^2} \ , \quad I_{2}=\frac{2H-\omega_{1}^2-v_{1}^2-v_{2}^2}{\omega_{2}^2-\omega_{1}^2} \ .
\ee
We infer that, whenever the Hamiltonian of a deformation of the Neumann-Rosochatius system is conserved and the two squared frequencies differ between them, one may define the Uhlenbeck constants 
through the above relation from the assumption that the two equalities (\ref{Redundancy}) still hold. If we make this assumption, the Hamiltonian (\ref{Hamiltonian}) leads to 
\begin{align}
\label{Uhlenbeck}
I_{1} & = r_{1}^2-\frac{1}{\omega_{2}^2-\omega_{1}^2}\left[\frac{\left(r_{1}^{\vphantom{'}}r_{2}'-r_{2}^{\vphantom{'}}r_{1}'\right)^2}{1+\kappa_{-}^2r_{1}^2+\kappa_{+}^2r_{2}^2} 
- \frac{\left(\kappa_{+}\omega_{1}-\kappa_{-}\omega_{2}\right)^2r_{1}^2r_{2}^2}{1+\kappa_{-}^2r_{1}^2+\kappa_{+}^{2}r_{2}^2} \right. \nonumber \\ 
& \left. + \, \frac{\left(1+\kappa_{-}^2\right)r_{2}^2}{r_{1}^2}v_{1}^2 + \frac{\left(1+\kappa_{+}^2\right)r_{1}^2}{r_{2}^2}v_{2}^2 - 2 \kappa_{-}\kappa_{+}v_{1}v_{2} \right] \ , \\ 
I_{2} & = r_{2}^2+\frac{1}{\omega_{2}^2-\omega_{1}^2}\left[\frac{\left(r_{1}^{\vphantom{'}}r_{2}'-r_{2}^{\vphantom{'}}r_{1}'\right)^2}{1+\kappa_{-}^2r_{1}^2+\kappa_{+}^2r_{2}^2}
- \frac{\left(\kappa_{+}\omega_{1}-\kappa_{-}\omega_{2}\right)^2r_{1}^2r_{2}^2}{1+\kappa_{-}^2r_{1}^2+\kappa_{+}^{2}r_{2}^2} \right. \nonumber \\
& \left. + \, \frac{\left(1+\kappa_{-}^2\right)r_{2}^2}{r_{1}^2}v_{1}^2 + \frac{\left(1+\kappa_{+}^2\right)r_{1}^2}{r_{2}^2}v_{2}^2 
- 2 \kappa_{-}\kappa_{+}v_{1}v_{2} \right] \ , 
\end{align}  
where we have used that the constraint (\ref{Constraint}) implies that $\left(r_{1}^{\vphantom{'}}r_{2}'-r_{2}^{\vphantom{'}}r_{1}'\right)^2 =r_{1}'^2+r'^{2}_{2}$. 

%%%%%%%%%%%%%%%%%%%%%%%%%%%%%%%%%%%%%%%%%%%%%%%%%%%%%%%%%%%%%%%%%%%%%%%%
%%%%%%%%%%%%%%%%%%%%%%%%%%%%%%%%%%%%%%%%%%%%%%%%%%%%%%%%%%%%%%%%%%%%%%%%

\section{Spinning string solutions}

\label{Tres}

In this section we will construct the various classes of elliptic solutions to the two-parameter deformation of the Neumann-Rosochatius system corresponding to closed strings rotating in~$\Manifold$.
In order to proceed, we will introduce the parameterization of $r_{a}^2$ in terms of the ellipsoidal coordinate $\zeta$, defined by
\be
\frac{r_{1}^2}{\zeta-\omega_{1}^2}+\frac{r_{2}^2}{\zeta-\omega_{2}^2}=0 \ ,
\ee
or, equivalently, by
\be
\label{Unequal}
r_{1}^2=\frac{\zeta-\omega_{1}^2}{\omega_{2}^2-\omega_{1}^2} \ , \quad r_{2}^2=\frac{\omega_{2}^2-\zeta}{\omega_{2}^{2}-\omega_{1}^2} \ .
\ee
The ellipsoidal coordinate satisfies
\be
-\frac{\zeta'^2}{4\left(\zeta-\omega_{2}^2\right)\left(\zeta-\omega_{1}^2\right)}=(r_{1}^{\vphantom{'}}r_{2}'-r_{2}^{\vphantom{'}}r_{1}')^2 \ .
\ee
Assuming that $\omega_{1}^2 < \omega_{2}^2$ without loss of generality, the conditions $0 \le r_{a}^2 \le 1$ entail that $\omega_{1}^2 \le \zeta \le \omega_{2}^2$. 
If we write, for instance, the first Uhlenbeck constant (\ref{Uhlenbeck}) in terms of the ellipsoidal coordinate, we find that
\be
\label{Differential}
\zeta'^2 = 4P_{4} \left(\zeta\right) \ ,
\ee
where $P_{4}(\zeta)$ is the fourth order polynomial 
\be
\label{Quartic}
\begin{split}
P_{4} \left( \zeta \right) & = \frac{\omega_{2}^2 - \omega_{1}^2+\kappa_{-}^2\left(\zeta-\omega_{1}^2\right)-\kappa_{+}^2\left(\zeta-\omega_{2}^2\right)}{\omega_{2}^2-\omega_{1}^2}
\big[ \left(\omega_{2}^2 - \omega_{1}^2\right) I_{1} \left( \zeta - \omega_{1}^2 \right) \left( \zeta - \omega_{2}^2 \right)  \\ 
& - \left( \zeta - \omega_{1}^2 \right)^{2} \left(\zeta- \omega_{2}^2 \right) - \left(1+\kappa_{+}^2\right)\left(\zeta- \omega_{2}^2 \right)^2v_{1}^2-\left(1+\kappa_{-}^2\right)\left(\zeta-\omega_{1}^2\right)^2v_{2}^2 \\ 
& - 2 \kappa_{-}\kappa_{+}v_{1}v_{2} \left( \zeta- \omega_{1}^2\right) \left(\zeta-\omega_{2}^2\right) \big] 
+ \frac{\left(\kappa_{+}\omega_{1}-\kappa_{-}\omega_{2}\right)^2\left(\zeta - \omega_{1}^2 \right)^2 \left(\zeta-\omega_{2}^2\right)^2}{\left(\omega_{2}^2-\omega_{1}^2\right)^2} \ .
\end{split}
\ee
Even though the expressions for the roots are too lengthy to be considered explicitly, the ordinary differential equation could still be solved formally.~\footnote{The usage of the second Virasoro constraint in (\ref{Virasoro}) does not introduce any appreciable simplification.} 
We must first discriminate between the two alternative domains for the parameters stated below equation~(\ref{cirteM}), namely, the case of real deformation parameters 
and the one in which they are purely imaginary. The polynomial (\ref{Quartic}) is actually quartic if $\left|\kappa_{-}\omega_{1}-\kappa_{+}\omega_{2}\right|>0$, whereas it is cubic when $\kappa_{-}\omega_{1}-\kappa_{+}\omega_{2}=0$. If the former inequality is satisfied, we need to further differentiate three cases for both real and purely imaginary deformation parameters. Those are the case in which all roots are real, the one that involves two real roots and two complex-conjugated roots, and, lastly, that where there are two pairs of complex-conjugated roots. In principle, the three possibilities could emerge as the discriminant of the quartic polynomial has no definite sign for general values of the parameters satisfying the second Virasoro constraint in (\ref{Virasoro}). The undeformed limit entails the divergence of one of the four roots. If we have $\kappa_{-}\omega_{1}-\kappa_{+}\omega_{2}=0$ instead,  the discriminant has no definite sign either, and the polynomial may possess three real roots, or one real root and two-complex conjugated roots. None of the roots diverges in the undeformed limit.

Moreover, we emphasize here that the differential equation (\ref{Differential}) may be used to determine the angular momenta in terms of the roots of the polynomial, 
which potentially leads to the energy as a function of the latter momenta drawing on the Virasoro contraints. However, the presence of deformation parameter ravels the integrands as displayed in (\ref{Conserved}), 
which leads to intricate expressions for the angular momenta that we will not present here. 
 
Before we set out with the analysis of the solutions to~(\ref{Differential}), it is convenient to stress some aspects of the undeformed limit of $P_{4}(\zeta)$. If the deformation parameters $\kappa_{\pm}$ are set to zero in~(\ref{Quartic}), 
it becomes the third-order polynomial 
\be
\begin{split}
\label{Cubic}
P_{3} \left(\zeta\right) &= \left(\omega_{2}^2-\omega_{1}^2\right)I_{1}\left(\zeta- \omega_{2}^2\right)\left(\zeta-\omega_{1}^2\right) +
\left(\zeta-\omega_{1}^2\right) ^2\left( \omega_{2}^2 - \zeta \right)  \\ &- \left(\zeta-\omega_{1}^2\right)^2 v_{2}^{2} - \left(\zeta- \omega_{2}^2\right)^2v_{1}^2 \ .
\end{split}
\ee
The discriminant of this polynomial has no definite sign for general values of the parameters satisfying the second Virasoro constraint of (\ref{Virasoro}), and hence $P_{3}(\zeta)$ may have three real roots or one real root together with a pair of complex-conjugated roots. If we denote the roots by $\bar{\zeta}_{i}$, with $i=1,2,3$, we can express the equation (\ref{Differential}) in the undeformed limit like
\be
\label{CubicRoots}
\zeta'^2=-4\overset{3}{\underset{i=1}{\prod}} \left(\zeta-\bar{\zeta}_{i}\right) \ .
\ee
We stress that the coefficient of the third-order term equals minus four (minus one in $P_{3}\left(\zeta\right)$) since it will turn out to be important regarding the study of the undeformed limit. 

Let us briefly review the type of solutions to the equation above that we can construct. First and foremost, they depend on the reality conditions of the roots $\bar{\zeta}_{i}$. The solutions are also constrained by the reality of the right-hand side of (\ref{CubicRoots}) and the requirement of boundedness $\omega_{1}^2\leq\zeta\leq\omega_{2}^2$. For instance, let us consider that the three roots $\bar{\zeta}_{i}$ are real and different, which we can take to be ordered in the hierarchy $\bar{\zeta}_{1}<\bar{\zeta}_{2}<\bar{\zeta}_{3}$. The simplest solutions that we can construct are constant-radii solutions with $\zeta=\bar{\zeta_{i}}$, for which we should have $\omega_{1}^2\leq\bar{\zeta}_{i}\leq\omega_{2}^2$. Apart from them, we can also construct periodic rotating solutions with non-constant radii. Boundedness in this case requires the ellipsoidal coordinate to be confined to the interval $\bar{\zeta}_{2}\leq\zeta\leq\bar{\zeta}_{3}$, whose bounds should likewise satisfy $\omega_{1}^2\leq\bar{\zeta}_{2}<\bar{\zeta}_{3}\leq\omega_{2}^2$. We can also focus on the degenerate limit in which two roots coalesce. If $\bar{\zeta}_{1}=\bar{\zeta}_{2}$, we have a giant-magnon solution that constitutes an aperiodic limit of the rotating solution. If $\bar{\zeta}_{2}=\bar{\zeta}_{3}$, we have a constant-radii solution, since the interval of $\zeta$ collapses to a point. If $\bar{\zeta}_{1}=\bar{\zeta}_{2}=\bar{\zeta}_{3}$ is further satisfied, we are always led to this last case. Besides, if two roots are non-real and complex-conjugated to each other, we need to set the ellipsoidal coordinate to the real root, leading us to a constant-radii solution. 

It is worthwhile to bear the scheme of the undeformed scenario in mind when the problem of constructing solutions to (\ref{Differential}) is addressed, since an analogous analysis will be deployed therein. In addition, such a scheme lies the foundations for the study of the extension of the solutions under the presence of the deformation.

\subsection{Real deformation} 

\label{Real}

We will study now the case where the deformation parameters $\kappa_{\pm}$ are real. We will first consider the scenario in which all the roots are real, and denote them by $\zeta_{\alpha}$, with $\alpha=1,2,3,4$. Let us further require the condition $\kappa_{-}\omega_{1}-\kappa_{+}\omega_{2}\neq0$ that ensures that the polynomial (\ref{Quartic}) is indeed quartic. Accordingly, it is possible to express
\be
\label{Equation}
\zeta'^2=4\frac{\left(\kappa_{-}\omega_{1}-\kappa_{+}\omega_{2}\right)^2}{\left(\omega_{2}^2-\omega_{1}^2\right)^2} \overset{4}{\underset{\alpha=1}{\prod}} \left(\zeta-\zeta_{\alpha}\right) \ .
\ee
In the non-degenerate case, where all the roots are different, we may choose the hierarchy $\zeta_{1}<\zeta_{2}<\zeta_{3}<\zeta_{4}$ without loss of generality. Firstly, we look out for general rotating solutions, and hence we assume the non-trivial dependence of $\zeta$ on the spatial world-sheet coordinate. The ordinary differential equation can then be solved in terms of a
direct integration and the subsequent inversion of the expression thus obtained. But before integrating we need to find out the legitimate domain of the ellipsoidal coordinate. 
In order for the square root of the right-hand side of the equation (\ref{Equation}) to be real, we should have either $\zeta\le\zeta_{1}$, $\zeta_{2}\le\zeta\le\zeta_{3}$ or $\zeta_{4}\le\zeta$. 
The first and the last possibilities lead to unbounded solutions that are incompatible with the condition $\omega_{1}^2\le\zeta\le\omega_{2}^2$, 
since the derivative of~(\ref{Equation}) cannot attain the two critical points necessary for the ellipsoidal coordinate to be confined to such an interval (this fact may also be shown explicitly proceeding as below). 
Therefore, the range is $\zeta_{2}\le\zeta\le\zeta_{3}$. We should perform the following integration,
\be
\label{Integration}
\begin{split}
& 2 \frac{\abs{\kappa_{-}\omega_{1}-\kappa_{+}\omega_{2}}}{\omega_{2}^2-\omega_{1}^2}\left(\sigma-\sigma_{0}\right)=\int_{\zeta_{2}}^{\zeta}\frac{\dif\eta}{\sqrt{\left(\zeta_{4}-\eta\right)\left(\zeta_{3}-\eta\right)\left(\eta-\zeta_{2}\right)\left(\eta-\zeta_{1}\right)}} \\ 
& = \frac{2}{\sqrt{\left(\zeta_{4}-\zeta_{2}\right)\left(\zeta_{3}-\zeta_{1}\right)}}F\left(\arcsin_{\left[0,\pi/2\right]}\sqrt{\frac{\left(\zeta_{3}-\zeta_{1}\right)\left(\zeta-\zeta_{2}\right)}{\left(\zeta_{3}-\zeta_{2}\right)\left(\zeta-\zeta_{1}\right)}},\mu\right) \ ,
\end{split}
\ee
where $\sigma_{0}$ is an integration constant, $F(z,m)$ denotes the incomplete elliptic integral of the first kind, and the elliptic modulus $\mu$ is given by
\be
\label{Modulus}
\mu = \sqrt{\frac{\left(\zeta_{4}-\zeta_{1}\right)\left(\zeta_{3}-\zeta_{2}\right)}{\left(\zeta_{4}-\zeta_{2}\right)\left(\zeta_{3}-\zeta_{1}\right)}} \ .
\ee
Note that the elliptic modulus belongs to the fundamental domain since the complementary modulus does. We will we set to zero the integration constant $\sigma_{0}$, addressing the cases in which plays a relevant role separately. If we invert the equality, we obtain 
\be
\label{First}
\zeta=\zeta_{2}+\frac{\left(\zeta_{3}-\zeta_{2}\right)\left(\zeta_{2}-\zeta_{1}\right)\sn^2\left(\omega\sigma,\mu\right)}{\zeta_{3}-\zeta_{1}-\left(\zeta_{3}-\zeta_{2}\right)\sn^2\left(\omega\sigma,\mu\right)} \ ,
\ee
where we have defined the frequency
\be
\label{Periodicity}
\omega = \frac{\abs{\kappa_{-}\omega_{1}-\kappa_{+}\omega_{2}}}{\omega_{2}^2-\omega_{1}^2}\sqrt{\left(\zeta_{4}-\zeta_{2}\right)\left(\zeta_{3}-\zeta_{1}\right)} \ ,
\ee
and $\sn\left(z,m\right)$ denotes the Jacobian elliptic sine. The periodic boundary conditions for $r_{a}$ in (\ref{Parallel}) are inherited by the ellipsoidal coordinate, that must satisfy $\zeta\left(\sigma\right)=\zeta\left(\sigma+2\pi\right)$. Such a requirement entails that the frequency satisfies
\be
\label{Relationship}
\omega = \frac{n K\left(\mu\right)}{\pi} \ , 
\ee
where $K(m)$ denotes the complete elliptic integral of the first kind and $n$ is an integer number. Moreover, since $\omega_{1}^2\le\zeta\le\omega_{2}^2$, we have $\omega_{1}^2 \le \zeta_{2} < \zeta_{3} \le \omega_{2}^2$. Besides this solution, the equation (\ref{Equation}) also admits constant-radii solution  whose ellipsoidal coordinate equals one root. They are allowed whenever $\omega_{1}^2 \le \zeta_{\alpha} \le \omega_{2}^2$, likewise those of non-constant radii.

We may now study the solutions in which two or more roots of $P_{4}(\zeta)$ coalesce through the analysis of the degenerate limits of the general solution above. According to the ordering of roots that we have considered, the first set of cases to consider is $\zeta_{1}=\zeta_{2}$, $\zeta_{2}=\zeta_{3}$ and $\zeta_{3}=\zeta_{4}$. 
From the identification of the two degenerate roots in the expression of $\omega$, we conclude that all of them yield finite non-vanishing values for it. However, (\ref{Relationship}) does not hold anymore, since the coalescence of two roots implies the degeneration of the periodicity properties of the solution, as we will show below.

Let us first assume that $\zeta_{3}=\zeta_{4}$. Since the elliptic modulus $\mu$ becomes one, the general solution reduces to
\be
\label{GM}
\zeta=\zeta_{2}+\frac{\left(\zeta_{3}-\zeta_{2}\right)\left(\zeta_{2}-\zeta_{1}\right)\tanh^2\left(\omega\sigma\right)}{\zeta_{3}-\zeta_{1}-\left(\zeta_{3}-\zeta_{2}\right)\tanh^2\left(\omega\sigma\right)} \ . 
\ee
The degeneration is reflected as an aperiodic limit of $\zeta$. The period becomes infinite and thus~(\ref{Relationship}) does not hold. We may also deduce the aperiodicity from the latter relationship, which diverges when $\mu=1$. The solution then represents a giant-magnon configuration with non-trivial real deformation parameters. From the target-space viewpoint, we can depict the situation as follows. In general, (\ref{First}) encodes the polar angle $\theta\left(\sigma\right)$ of the embedding of the world-sheet into the two-parameter deformation of the three-sphere $\Manifold$. It exhibits the periodic boundary conditions $\theta\left(\sigma\right)=\theta\left(\sigma+2\pi\right)$, which are governed by the frequency satisfying (\ref{Relationship}). Due to this relation, the more $\zeta_{3}$ and $\zeta_{4}$ approach each other, the more distance should range the argument of $\sn\left(z,m\right)$ to bring $\theta\left(\sigma\right)$ until $\theta\left(\pi\right)$ and then bring it back to the initial point $\theta\left(0\right)$. In the limit where $\zeta_{3}$ and $\zeta_{4}$ merge, the Jacobian elliptic sine loses its periodicity properties and becomes a hyperbolic tangent. The polar angle accordingly is unable to go back to $\theta\left(0\right)$ as it just reaches $\theta\left(\pi\right)$ asymptotically. Of course, this picture is not exclusive of (\ref{GM}) and carries over into the giant-magnon solutions of the undeformed setting. The distinguished property of the present framework rather consists in allowing us to construct two different giant-magnon solutions instead of just one.

Indeed, the solution (\ref{GM}) is not the only giant-magnon configuration that can emerge. The degenerate limit $\zeta_{1}=\zeta_{2}$ leads us again to an infinite period in view of~(\ref{Relationship}). However, if we take this limit directly in (\ref{First}), we obtain the constant solution $\zeta=\zeta_{2}$. The apparent contradiction is a consequence of the choice of the integration constant in the general solution, which matters in the aperiodic limit. We can obtain an expression of the giant-magnon type for the ellipsoidal coordinate if we set $\sigma_{0}$ to an appropriate value prior to the application of the limit. In particular, if instead of zero we set it to be proportional to a quarter of a period of the elliptic function,
\be
\sigma_{0}=\frac{K\left(\mu\right)}{\omega}=\frac{\left(\omega_{2}^2-\omega_{1}^2\right)K\left(\mu\right)}{\abs{\kappa_{-}\omega_{1}-\kappa_{+}\omega_{2}}\sqrt{\left(\zeta_{4}-\zeta_{2}\right)\left(\zeta_{3}-\zeta_{1}\right)}} \ ,
\ee
we are led to
\be
\label{New}
\zeta=\zeta_{3}-\frac{\left(\zeta_{4}-\zeta_{3}\right)\left(\zeta_{3}-\zeta_{2}\right)\sn^2\left(\omega\sigma,\mu\right)}{\zeta_{4}-\zeta_{2}-\left(\zeta_{3}-\zeta_{2}\right)\sn^2\left(\omega\sigma,\mu\right)} \ .
\ee
To arrive to this expression we have employed the property 
\be
\sn^2 \left(z - K\left(m\right),m\right)=\frac{\cn^2\left(z,m\right)}{1-m^2\sn^2\left(z,m\right)} \ ,
\ee
and $\cn\left(z,m\right)$ denotes the Jacobian elliptic cosine. This expression allows us to take the degenerate limit $\zeta_{1}=\zeta_{2}$ in (\ref{New}), leading us to
\be
\label{GMII}
\zeta=\zeta_{3}-\frac{\left(\zeta_{4}-\zeta_{3}\right)\left(\zeta_{3}-\zeta_{2}\right)\tanh^2\left(\omega\sigma\right)}{\zeta_{4}-\zeta_{2}-\left(\zeta_{3}-\zeta_{2}\right)\tanh^2\left(\omega\sigma\right)} \ ,
\ee
Therefore, the degeneration $\zeta_{1}=\zeta_{2}$ reduces the solution to a giant-magnon configuration. Notice that if we set $\zeta_{3}=\zeta_{4}$ in (\ref{New}), the ellipsoidal coordinate becomes $\zeta=\zeta_{3}$. The situation is complementary to (\ref{First}).

The remaining degenerate limit $\zeta_{2}=\zeta_{3}$ yields to the constant radii solution $\zeta=\zeta_{2}$. The elliptic modulus vanishes and the relationship (\ref{Relationship}) breaks down because the period of a constant is arbitrary. A target-space picture may be portrayed along the lines of what we have previously considered for the giant-magnon solution. Since the general solution satisfies $\zeta_{2}\leq\zeta\leq\zeta_{3}$, the polar angle $\theta(\sigma)$ oscillates between the pair of points $\theta\left(0\right)$ and $\theta\left(\pi\right)$, which get closer as $\zeta_{2}$ and $\zeta_{3}$ get closer. In the limit $\zeta_{2}=\zeta_{3}$, the two points $\theta(0)$ and $\theta\left(\pi\right)$ merge and $\theta\left(\sigma\right)$ collapses to a point in $\Manifold$. Actually, we can merely construct constant-radii solutions when more than two roots degenerate. If three consecutive roots coalesce, the hierarchy of roots $\zeta_{1}<\zeta_{2}<\zeta_{3}<\zeta_{4}$ entails that $\zeta$ unavoidably collapses to a point. The statement also applies to the giant-magnon solutions (\ref{GM}) and (\ref{GMII}) that, despite having lost the periodicity properties, still range between $\zeta_{2}$ and $\zeta_{3}$. If we have $\zeta_{1}=\zeta_{2}<\zeta_{3}=\zeta_{4}$, we need to study the limit in terms of the giant-magnon solutions. We conclude that the functional dependence is erased due to the vanishing of the prefactor, so the solution becomes equal to $\zeta=\zeta_{2}$ in (\ref{GM}) and $\zeta=\zeta_{3}$ in (\ref{GMII}). However, if we had started with an unsuited choice of integration constant, we would have obtained $\zeta=\zeta_{3}$ in (\ref{GM}) and $\zeta=\zeta_{2}$ in (\ref{GMII}). We can puzzle this result out noting that in general the solution ranges between the endpoints $\zeta_{2}$ and $\zeta_{3}$, or equivalently $\theta\left(0\right)$ and $\theta\left(\pi\right)$. In the limit in which both endpoints correspond to a degenerate root, the solution disintegrates and localizes on them. Note that such a disintegration requires the presence of four real roots undergoing a double pairwise degeneration. Finally, if the four roots coalesce, we have that the solution equals the available root.

Let us turn to the undeformed elliptic counterpart of the solution (\ref{First}). In principle, it depends on the root $\zeta_{\alpha}$ of the quartic polynomial that diverges in the limit of vanishing deformation parameters. In order to proceed we should take into account two facts. Firstly, we have that for $\kappa_{\pm}$ close enough to zero the hierarchy $\zeta_{1}<\zeta_{2}<\zeta_{3}<\zeta_{4}$ has to be respected. Secondly, in the non-degenerate case, taking the limit in which $\kappa_{\pm}$ tend to zero cannot turn two real roots into a pair of complex-conjugated ones, and vice versa, if the remaining parameters are fixed. Therefore, there are two possibilities that may permit us to recover the cubic polynomial $P_{3}(\zeta)$ as expressed in (\ref{CubicRoots}), with $\bar{\zeta}_{i}$ denoting three real roots assumed to be ordered like $\bar{\zeta}_{1}<\bar{\zeta}_{2}<\bar{\zeta}_{3}$. One alternative consists in the divergence of the greatest root as $\zeta_{4}\rightarrow\infty$ and the other consists in the divergence of the lowest one as $\zeta_{1}\rightarrow-\infty$. However, the occurrence of the latter possibility can be excluded drawing on the expression of the cubic polynomial $P_{3}\left(\zeta\right)$ in (\ref{CubicRoots}), whose third-order coefficient is minus one. If the limit $\zeta_{1}\rightarrow-\infty$ occurred, we would recover a third-order coefficient equal to plus one. The undeformed limit should be then realized by 
\be
\label{One}
\zeta_{4} \rightarrow \infty \ , \quad \zeta_{i} \rightarrow \bar{\zeta}_{i} \ , 
\quad \with \quad \frac{\left(\kappa_{-}\omega_{1}-\kappa_{+}\omega_{2}\right)^2}{\left(\omega_{2}^2-\omega_{1}^2\right)^2}\zeta_{4}\rightarrow 1 \ .
\ee
If we apply this limit in (\ref{Modulus}) and (\ref{Periodicity}) , we find
\be
\omega \rightarrow \bar{\omega} = \sqrt{\bar{\zeta}_{3}-\bar{\zeta}_{1}} \ , \quad \mu\rightarrow \bar{\mu}=\sqrt{\frac{\bar{\zeta}_{3}-\bar{\zeta}_{2}}{\bar{\zeta}_{3}-\bar{\zeta}_{1}}} \ .
\label{omegalimit}
\ee
Regarding the general solution (\ref{First}), the application of the limit of vanishing deformation parameters thus amounts to the proper replacement of unbarred by barred parameters. The periodicity condition (\ref{Relationship}) also holds once the quantities are barred. The undeformed solution can be simplified to obtain the expressions for $\zeta$ considered in \cite{NR,NR2}. Either shifting the spatial world-sheet coordinate as $\sigma \mapsto \sigma + K \left(\bar{\mu}\right)/\bar{\omega}$ or, equivalently, taking the undeformed limit in the alternative expression (\ref{New}), leads us to
\be
\label{Uno}
\zeta \rightarrow \bar{\zeta}_{3} - \left(\bar{\zeta}_{3} - \bar{\zeta}_{2}\right) \sn^2\left(\bar{\omega}\sigma,\bar{\mu}\right) \ .
\ee
It is worth to point out that the requirement of boundedness of the ellipsoidal coordinate $\omega_{1}^2\leq\zeta_{2}<\zeta_{3}\leq\omega_{2}^2$ carries over into an analogous condition in the undeformed limit, namely, $\omega_{1}^2\leq\bar{\zeta}_{2}<\bar{\zeta}_{3}\leq\omega_{2}^2$. Moreover, the constant-radii solutions that are present in the non-degenerate scenario with $\zeta=\zeta_{i}$ (but not with $\zeta=\zeta_{4}$) reduce to their undeformed counterparts $\zeta=\bar{\zeta}_{i}$.

We may acquire insight into the behavior of the solution by means of the study of the connection between the degenerate limits of the deformed and undeformed settings. Let us consider first the degenerate limits with $\zeta_{1}=\zeta_{2}$ and $\zeta_{2}=\zeta_{3}$. When $\zeta_{1}=\zeta_{2}$ we have obtained the giant-magnon solution (\ref{GMII}). We can argue that the solution is the double Yang-Baxter deformation with real deformation parameters of the ordinary giant-magnon solution. Indeed, taking the limit (\ref{One}) in the solution (\ref{GMII}) provides us
\be
\label{GMIII}
\zeta \rightarrow \bar{\zeta}_{3} - \left(\bar{\zeta}_{3} - \bar{\zeta}_{2}\right) \tanh^2\left(\bar{\omega}\sigma\right) \ .
\ee
This expression is obtained again from (\ref{Uno}) in the degenerate limit where $\bar{\zeta}_{1}=\bar{\zeta}_{2}$. It is an aperiodic limit of infinite period, because (\ref{omegalimit}) states that elliptic modulus $\bar{\mu}$ equals one in this case. Similarly, we can argue that the degenerate limit  with $\zeta_{2}=\zeta_{3}$ (and also with $\zeta_{1}=\zeta_{2}=\zeta_{3}$) leading to $\zeta=\zeta_{2}$ is related to $\zeta=\bar{\zeta}_{2}$ in the degenerate limit with $\bar{\zeta}_{2}=\bar{\zeta}_{3}$ (respectively $\bar{\zeta}_{1}=\bar{\zeta}_{2}=\bar{\zeta}_{3}$). 

The behavior of the giant-magnon solution (\ref{GM}), for which $\zeta_{3}=\zeta_{4}$, is more subtle. The direct application of the undeformed limit (\ref{One}) therein leads us to an unbounded solution. However, this is not an actual solution in the undeformed limit, since the divergence of $\zeta_{4}$ would imply that the root $\zeta_{3}$ also diverges. This possibility is denied by the fact that $P_{3}\left(\zeta\right)$ is a cubic polynomial irrespective of the value of the parameters in the solution. In contradistinction to the two the cases with $\zeta_{1}=\zeta_{2}$ and $\zeta_{2}=\zeta_{3}$, the equality $\zeta_{3}=\zeta_{4}$ cannot hold in the undeformed limit. In fact, the solution (\ref{GM}) ceases to exist before the undeformed limit is reached even if the equality $\zeta_{3}=\zeta_{4}$ remains valid. To clarify why this is the case, we find useful to resort to the viewpoint of the target-space embedding of the world-sheet. From this perspective, the reason underlying the disappearance of the solution is the following. The initial deformed scenario involves the partially degenerate hierarchy $\zeta_{1}<\zeta_{2}<\zeta_{3}=\zeta_{4}$, whose associated solution (\ref{GM}) represents a polar angle $\theta\left(\sigma\right)$ starting at $\theta(0)$ when $\zeta=\zeta_{2}$ and ending at $\theta\left(\pi\right)$ at $\zeta=\zeta_{3}$. The more $\kappa_{\pm}$ approach zero, the more the degenerate roots $\zeta_{3}=\zeta_{4}$ increase. If the equality $\zeta_{3}=\zeta_{4}$ is assumed to hold (for instance, tuning the parameters of the solution), beyond a small enough value of $\kappa_{\pm}$ the bound $\zeta_{3}\leq\omega_{2}^2$ cannot be satisfied. The violation of the bound entails that a real solution for $\theta\left(\sigma\right)$ ceases to exist because of the complexification of $\theta\left(\pi\right)$, which renders the associated solution unacceptable. Therefore, we find that the presence of the real deformation parameters gives rise to a giant-magnon solution that lacks any counterpart in the undeformed limit. Following an analogous reasoning, we conclude that the deformation parameters also yield constant-radii solutions that have no correlatives in the undeformed setting. Such solutions are those involving $\zeta=\zeta_{4}$, which emerge when either none, three or all roots degenerate.

We consider now the case where the $P_{4} \left(\zeta\right)$ involves two real and two complex-conjugated roots. If two of the four roots are complex-conjugated to each other, the positivity of the right-hand side of (\ref{Differential}) implies that just unbounded solutions could be obtained for $\zeta$ unless it equals one of the two real roots. The statement is valid in either the non-degenerate and degenerate cases. In a non-degenerate scenario, the undeformed setting involving one real root and two complex-conjugated roots is retrieved when the greatest real root diverges. Thus, if $\zeta$ equals the greatest real root, it lacks any undeformed counterpart, whereas it reduces to its undeformed analogue if it equals the lowest real root. The degenerate limit involving the identification of two roots of $P_{4}\left(\zeta\right)$ in this case is either realized by the equality of the two real roots or by the equality of the two complex-conjugated roots, which become real. The former possibility possesses a constant-radii solution that lacks any counterpart since the undeformed polynomial $P_{3}\left(\zeta\right)$ cannot be directly recovered. On the contrary, the latter is comprised by the previous analysis on the degenerate limit involving two roots of a hierarchy of four real roots, and hence it may have an undeformed counterpart. The cases with three or four degenerate roots are also encompassed in this analysis. Moreover, if $P_{4}\left(\zeta\right)$ involves two pairs of complex-conjugated roots, (\ref{Equation}) neither admits any solution for $\zeta$ nor is connected with $P_{3}\left(\zeta\right)$ through the divergence of one of the roots. In order to obtain acceptable solutions we need to focus on the degenerate limits that have been previously discussed.

Let us assume now that $\kappa_{-}\omega_{1}-\kappa_{+}\omega_{2}=0$. We will exclude the possibility where either $\kappa_{-}$ or $\kappa_{+}$ vanish, since we would be addressing the construction of solutions to a truncation of the $\eta$-deformation of the Neumann-Rosochatius system. This problem has been already studied in \cite{NReta}, and we refer the reader to that reference. If $\kappa_{-}\omega_{1}-\kappa_{+}\omega_{2}=0$, the polynomial in $P_{4}\left(\zeta\right)$ lowers its degree by one and we are led to a cubic polynomial $Q_{3}\left(\zeta\right)$. Let us denote its roots by $\zeta_{i}$, with $i=1,2,3$. Its discriminant has not definite sign for parameters satisfying the second Virasoro constraint of (\ref{Virasoro}). Hence, the polynomial $Q_{3}\left(\zeta\right)$ may present three real roots, or one real root and two complex-conjugated ones. It could be formally rephrased alike $P_{3}\left(\zeta\right)$ in the equation (\ref{CubicRoots}) if the roots $\bar{\zeta}_{i}$ are replaced by $\zeta_{i}$ and the coefficient of the third-order term that equals minus one is extended to $A$. Regarding the latter, instead of writting down its expression directly, it is convenient to employ the available constraints to eliminate some parameters and simplify the resultant expression. First of all, $\kappa_{-}\omega_{1}-\kappa_{+}\omega_{2}=0$ allows us to express one of the frequencies $\omega_{a}$ in terms of the other and $\kappa_{\pm}$. Notice that the condition $\omega_{1}^2\neq\omega_{2}^2$ required for the definition of the ellipsoidal coordinate entails that $\kappa_{\pm}^2$ must differ between them 
and that none $\omega_{a}$ can vanish. We underline the contrast with the condition needed when $\kappa_{-}\omega_{1}-\kappa_{+}\omega_{2}\neq0$, that is, the condition stating that the $\omega_{a}$ are different (while the smaller could vanish). We may then take $\omega_{2}=\kappa_{-}\omega_{1}/\kappa_{+}$. Our choice $\omega_{1}^2<\omega_{2}^2$ renders into $\kappa_{+}^2<\kappa_{-}^2$. We can then use the equation for $\omega_{2}$ in the second Virasoro constraint of (\ref{Virasoro}), thereby obtaining $v_{2}=-\kappa_{+}v_{1}/\kappa_{-}$. Following these steps, we find that the coefficient of the third-order term $A$ in $Q_{3}\left(\zeta\right)$ reads
\be
\label{A}
A=-1+\kappa_{-}^2-2\frac{\kappa_{-}\kappa_{+}}{\omega_{1}}+I_{1}\left(\kappa_{-}^2-\kappa_{+}^2\right)-\frac{\kappa_{+}^2\left(\kappa_{-}^2+\kappa_{+}^2\right)v_{1}^2}{\omega_{1}^2\kappa_{-}^2} \ .
\ee
We conclude that it has not definite sign for general values of the parameters involved. To obtain the undeformed limit in the explicit expression for the polynomial, we have to undo the transformations for $\omega_{2}$ and $v_{2}$ in $Q_{3}\left(\zeta\right)$ because they do not hold if $\kappa_{\pm}$ tend to zero. The limit does not involve the divergence of any root $\zeta_{i}$ because the degree of $Q_{3}\left(\zeta\right)$ is already three. In any case, it is manifest from (\ref{A}) that the undeformed limit provides minus one, as it should. 

The retrieval of $Q_{3}\left(\zeta\right)$ in terms of the roots of $P_{4}\left(\zeta\right)$ mimics the undeformed limit. Let us first assume that all the roots $\zeta_i$ are real and ordered in the hierarchy $\zeta_{1}<\zeta_{2}<\zeta_{3}$. Through a parallel reasoning to the one developed in the analysis of the undeformed limit, we conclude that the limit that takes $P_{4}\left(\zeta\right)$ to $Q_{3}\left(\zeta\right)$ depends on the sign of $A$. If $A<0$, the limit is supplied by (\ref{One}) once the proper replacements are performed. Namely, the substitution of $\bar{\zeta}_{i}$ by the roots $\zeta_{i}$, and the factor one in the rightmost condition by $A$. If $A>0$, we have instead that the lower root of $P_{4}\left(\zeta\right)$ should diverge as $\zeta_{1}\rightarrow-\infty$ in order for the hierarchy $\zeta_{1}<\zeta_{2}<\zeta_{3}<\zeta_{4}$ to be upheld and for the correct sign of $A$ to be recovered. The limit is then given by
\be
\label{Akin}
\zeta_{1} \rightarrow -\infty \ , \quad \zeta_{i+1} \rightarrow \zeta_{i} \ , 
\quad \with \quad \frac{\left(\kappa_{-}\omega_{1}-\kappa_{+}\omega_{2}\right)^2}{\left(\omega_{2}^2-\omega_{1}^2\right)^2}\zeta_{1}\rightarrow -A \ .
\ee
The left-hand side of the limit above refers to the roots of the quartic polynomial $P_{4}\left(\zeta\right)$, while its right-hand side refers to the roots of the cubic polynomial $Q_{3}\left(\zeta\right)$.
As far as the reduction of the solutions is concerned, they depend on the sign of $A$. If $A<0$, the solutions follow from those in undeformed limit if we take 
\be
\omega = \sqrt{-A\left(\zeta_{3}-\zeta_{1}\right)} \ , \quad \mu=\sqrt{\frac{\zeta_{3}-\zeta_{2}}{\zeta_{3}-\zeta_{1}}} \ .
\ee
We could then obtain the solutions for the case with $A>0$ through an analytically continuation of the expressions with $A<0$. By means of the properties of the Jacobian elliptic functions, we are eventually led to a similar solution that depends the complementary elliptic modulus of $\mu$. The result is consistent with the fact that the bound changes to $\zeta_{1}\leq\zeta\leq\zeta_{2}$, and so does the interval of the integration involved in the direct resolution of the equation $\zeta'^2=4Q_{3}\left(\zeta\right)$. The parallels with the undeformed scenario imply that the analysis of the degenerate limits of $P_{3}\left(\zeta\right)$ extends to those of $Q_{3}\left(\zeta\right)$. Conversely, the reduction of the deformed solutions (for which we should take $A<0$) to their undeformed counterparts is immediate. The degenerate and non-degenerate cases for $Q_{3}\left(\zeta\right)$ with one real root and a pair of complex-conjugated roots follow the same pattern as above. 

Lastly, it remains to address the point where $A=0$. It does not occur for particularly significant values of the parameters in (\ref{A}). Accordingly, the polynomial $Q_{3}\left(\zeta\right)$ lowers its degree to two. Irrespective of the sign of the discriminant of $Q_{3}\left(\zeta\right)$, the quadratic polynomial comes about as a consequence of the divergence of a real root. From the point of view of $P_{4}\left(\zeta\right)$, the situation is rather involved due to the various possibilities that could lead to this situation, but they always involve the simultaneous divergence of two roots. The techniques needed to address this case are formally the same as the ones that have been already employed. We will then omit an explicit treatment of this point for the sake of conciseness. The reader can consult the reduction of a solution involving three roots to a solution involving two roots in a different context in \cite{HNelliptic}.

\subsection{Purely imaginary deformation}

We will move now to the case of purely imaginary deformation parameters $\kappa_{\pm}$ whose moduli are less or equal than the unity (we will exclude from the analysis the case in which both moduli are equal to one). 
It is convenient to write $\kappa_{\pm}=i k_{\pm}$, where $k_{\pm}$ are real numbers whose absolute value do not equal the unity at the same time.
We will study first the scenario with $k_{-}\omega_{1}-k_{+}\omega_{2}\neq0$. Let us further assume that all the roots are real. If we denote them again by $\zeta_{\alpha}$, with $\alpha=1,2,3,4$, we may write
\be
\label{noitauqE}
\zeta'^2=-4\frac{\left(k_{-}\omega_{1}-k_{+}\omega_{2}\right)^2}{\left(\omega_{2}^2-\omega_{1}^2\right)^2} \overset{4}{\underset{\alpha=1}{\prod}} \left(\zeta-\zeta_{\alpha}\right) \ .
\ee 
Firstly, we will focus on general rotating solutions in the non-degenerate case, where we will consider the ordering $\zeta_{1}<\zeta_{2}<\zeta_{3}<\zeta_{4}$. We can then solve (\ref{noitauqE}) by direct integration. 
The reality of the square root of the right-hand side of (\ref{noitauqE}) and the boundedness of $\zeta$ require that either $\zeta_{1} \le \zeta \le \zeta_{2}$ or $\zeta_{3} \le \zeta \le \zeta_{4}$, but there is no additional property that implies the preference of one interval to the other.
Let us assume that $\zeta_{1} \le \zeta \le \zeta_{2}$. We can thus integrate the differential equation like
\be
\label{largetnI}
\begin{split}
&2\frac{\abs{k_{-}\omega_{1}-k_{+}\omega_{2}}}{\omega_{2}^2-\omega_{1}^2}\left(\sigma-\sigma_{0}\right)=\int_{\zeta_{1}}^{\zeta}\frac{\dif\eta}{\sqrt{\left(\zeta_{4}-\eta\right)\left(\zeta_{3}-\eta\right)\left(\zeta_{2}-\eta\right)\left(\eta-\zeta_{1}\right)}} \\ 
&=\frac{2}{\sqrt{\left(\zeta_{4}-\zeta_{2}\right)\left(\zeta_{3}-\zeta_{1}\right)}}F\left(\arcsin_{\left[0,\pi/2\right]}\sqrt{\frac{\left(\zeta_{4}-\zeta_{2}\right)\left(\zeta-\zeta_{1}\right)}{\left(\zeta_{2}-\zeta_{1}\right)\left(\zeta_{4}-\zeta\right)}},\mu\right) \ ,
\end{split}
\ee
where $\sigma_{0}$ is an integration constant and the elliptic modulus is
\be
\mu=\sqrt{\frac{\left(\zeta_{4}-\zeta_{3}\right)\left(\zeta_{2}-\zeta_{1}\right)}{\left(\zeta_{4}-\zeta_{2}\right)\left(\zeta_{3}-\zeta_{1}\right)}} \ . 
\label{mu}
\ee
If we set the integration constant to zero and invert this relation, we find 
\be
\label{Second}
\zeta=\zeta_{1}+\frac{\left(\zeta_{4}-\zeta_{1}\right)\left(\zeta_{2}-\zeta_{1}\right)\sn^2\left(\omega\sigma,\mu\right)}{\zeta_{4}-\zeta_{2}+\left(\zeta_{2}-\zeta_{1}\right)\sn^2\left(\omega\sigma,\mu\right)} \ ,
\ee
where the frequency $\omega$ is
\be
\omega = \frac{\abs{k_{-}\omega_{1}-k_{+}\omega_{2}}}{\omega_{2}^2-\omega_{1}^2} \sqrt{ \left(\zeta_{4}-\zeta_{2}\right)\left(\zeta_{3}-\zeta_{1}\right) } \ .
\label{omega}
\ee
If we choose instead the domain where $\zeta_{3}\le\zeta\le \zeta_{4}$, we obtain
\be
\label{Third}
\zeta = \zeta_{4}-\frac{\left(\zeta_{4}-\zeta_{1}\right) \left(\zeta_{4}-\zeta_{3}\right) \sn^2\left(\omega\sigma,\mu\right)}{\zeta_{3}-\zeta_{1}+\left(\zeta_{4}-\zeta_{3}\right)\sn^2\left(\omega\sigma,\mu\right)} \ ,
\ee
where $\mu$ and $\omega$ are as in equations (\ref{mu}) and (\ref{omega}) and the integration constant has been fixed to
\be
\sigma_{0}=\frac{K\left(\mu\right)}{\omega}=\frac{\left(\omega_{2}^2-\omega_{1}^2\right)K\left(\mu\right)}{\abs{k_{-}\omega_{1}-k_{+}\omega_{2}}\sqrt{\left(\zeta_{4}-\zeta_{2}\right)\left(\zeta_{3}-\zeta_{1}\right)}} \ .
\ee
Such a choice will facilitate the attainment of the giant-magnon solution limit of this solution. Of course, the complementary choice with $\sigma_{0}=0$ would have been also valid. Periodic boundary conditions imply that the frequency should satisfy a relationship analogous to (\ref{Relationship}) for the two solutions above,
\be
\label{Relation}
\omega = \frac{n K\left(\mu\right)}{\pi} \ . 
\ee
Moreover, the restriction $\omega_{1}^2 \le \zeta \le \omega_{2}^2$ implies that $\omega_{1}^2 \le \zeta_{1} < \zeta_{2} \le \omega_{2}^2$ in the former case, and that $\omega_{1}^2 \le \zeta_{3} < \zeta_{4} \le \omega_{2}^2$ in the latter. We finally note that, apart from the (\ref{Second}) and~(\ref{Third}), constant-radii solutions with $\zeta=\zeta_{\alpha}$ are also admissible whenever $\omega_{1}^2 \le\zeta_{\alpha} \le \omega_{2}^2$.

The degenerate cases for these two solutions are as follows. When two roots merge, the required hierarchy allows either $\zeta_{1}=\zeta_{2}$, $\zeta_{2}=\zeta_{3}$ or $\zeta_{3}=\zeta_{4}$. 
The coalescence of two roots leads to finite non-vanishing values for the frequency. As opposed to the degenerate limit of the regime where the deformation parameters are real, the equation (\ref{Relation}) holds for certain solutions. In the case where $\zeta_{1}=\zeta_{2}$, the elliptic modulus vanishes and the periodicity condition amounts to $\omega=n/2$. Regarding~(\ref{Second}), the range of the ellipsoidal coordinate collapses to a point and it reduces to $\zeta=\zeta_{1}$. In this case, (\ref{Relation}) breaks down as the period of a constant is arbitrary. On the contrary, the formula~(\ref{Third}) becomes
\be
\label{Limit}
\zeta = \zeta_{4}-\frac{\left(\zeta_{4}-\zeta_{1}\right) \left(\zeta_{4}-\zeta_{3}\right) \sin^2\left(\omega\sigma\right)}{\zeta_{3}-\zeta_{1}+\left(\zeta_{4}-\zeta_{3}\right)\sin^2\left(\omega\sigma\right)} \ ,
\ee
which is consistent with the fact that the range remains $\zeta_{3}\leq\zeta\leq\zeta_{4}$. The equation $\omega=n/2$ states nothing but the periodicity condition for trigonometric functions. If $\zeta_{2}=\zeta_{3}$, the relation (\ref{Relation}) does not hold since $\mu=1$ and hence the period tends to infinity. Accordingly, both (\ref{Second}) and (\ref{Third}) turn into giant-magnon solutions. The expression of the former becomes
\be
\label{GMIV}
\zeta = \zeta_{1}+\frac{\left(\zeta_{4}-\zeta_{1}\right)\left(\zeta_{2}-\zeta_{1}\right)\tanh^2\left(\omega\sigma\right)}{\zeta_{4}-\zeta_{2}+\left(\zeta_{2}-\zeta_{1}\right)\tanh^2\left(\omega\sigma\right)} \ ,
\ee
whereas that of the latter reads
\be
\label{GMV}
\zeta = \zeta_{4}-\frac{\left(\zeta_{4}-\zeta_{1}\right) \left(\zeta_{4}-\zeta_{2}\right) \tanh^2\left(\omega\sigma\right)}{\zeta_{2}-\zeta_{1}+\left(\zeta_{4}-\zeta_{2}\right)\tanh^2\left(\omega\sigma\right)} \ .
\ee
Analogously to the subsection \ref{Real}, we have then found that there exist two distinct giant-magnon solutions. Nonetheless, the two solutions may coexist as opposed to the case there. Finally, if $\zeta_{3}=\zeta_{4}$, we have again $\omega=n/2$. The solution (\ref{Second}) accordingly reads
\be
\label{timiL}
\zeta = \zeta_{1}+\frac{\left(\zeta_{3}-\zeta_{1}\right)\left(\zeta_{2}-\zeta_{1}\right)\sin^2\left(\omega\sigma\right)}{\zeta_{3}-\zeta_{2}+\left(\zeta_{2}-\zeta_{1}\right)\sin^2\left(\omega\sigma\right)} \ ,
\ee
while (\ref{Third}) becomes a constant-radii solution $\zeta=\zeta_{3}$ owing to the collapse of the corresponding interval. 

When three roots merge, the behavior of the solution again depends on the interval chosen for the ellipsoidal coordinate. If $\zeta_{1}=\zeta_{2}=\zeta_{3}$, the solution (\ref{Second}) becomes equal to the degenerate root since the interval for $\zeta$ contracts to a point. 
Regarding (\ref{Third}), we should consider either its trigonometric reduction (\ref{Limit}) or its associated giant-magnon solution (\ref{GMV}), and follow the reasoning presented in subsection \ref{Real} when four real roots undergo a double pairwise degeneration.
Proceeding in this way, we conclude that, starting from whichever degenerate limits, the solution disintegrates and becomes localized at its two endpoints $\zeta=\zeta_{3}$ and $\zeta=\zeta_{4}$. 
The roles of the two solutions are swapped if $\zeta_{2}=\zeta_{3}=\zeta_{4}$, where (\ref{Third}) equals the degenerate root and both the trigonometric and the aperiodic reduction of (\ref{Second}) disintegrates and localizes on its two endpoints. 
Moreover, if $\zeta_{1}=\zeta_{2}<\zeta_{3}=\zeta_{4}$, the admissible intervals of the ellipsoidal coordinate shrink to two points and we have that (\ref{Second}) becomes $\zeta=\zeta_{1}$ and (\ref{Third}) becomes $\zeta=\zeta_{3}$. If the four roots merge, we have that $\zeta=\zeta_{\alpha}$ for both solutions.

Let us turn to the undeformed limit of the two solutions in the non-degenerate setting. We are then led to study the reduction of the four real roots $\zeta_{\alpha}$ above to the corresponding three real roots of $P_{3}\left(\zeta\right)$. We denote them by $\bar{\zeta}_{i}$, with $i=1,2,3$, and we assume that they are ordered like $\bar{\zeta}_{1} < \bar{\zeta}_{2} < \bar{\zeta}_{3}$. The alternatives that respect the hierarchy $\zeta_{1} < \zeta_{2} < \zeta_{3} < \zeta_{4}$ are formally the same as the ones in the previous
subsection,
namely, the divergence of the greatest root $\zeta_{4}\rightarrow\infty$ and the divergence of the lowest root $\zeta_{1}\rightarrow-\infty$. The former is ruled out by the prefactor minus four in the expression (\ref{CubicRoots}), since this limits rather leads to plus four. Therefore, we need that
\be
\label{Two}
\zeta_{1}\rightarrow-\infty \ , \quad \zeta_{i+1} \rightarrow \bar{\zeta}_{i} \ , 
\quad \with \quad \frac{\left(k_{-}\omega_{1}-k_{+}\omega_{2}\right)^2}{\left(\omega_{2}^2-\omega_{1}^2\right)^2}\zeta_{1}\rightarrow -1 \ .
\ee
The preceding couple of distinct solutions are to be paired with their undeformed correlatives. Nonetheless, the solution (\ref{Second}) lacks such a counterpart as the root that plays the role of lower bound diverges. 
The absence of undeformed counterparts is shared by any reduction of both solutions involving the degeneration of $\zeta_{1}$, and also by the constant-radii solution $\zeta=\zeta_{1}$.
As a matter of fact, the solution (\ref{Second}) ceases to exist before the undeformed limit is reached, since at some point the bound $\omega_{1}^2\leq \zeta_{1} $ should be violated in order for (\ref{Two}) to occur. We have provided a description of this phenomenon in an analogous context in subsection \ref{Real}, and for that reason we omit it here. We content ourselves with emphasizing that the lack of an undeformed counterpart for the solution (\ref{Second}) points out the emergence of new classes of solutions due to the presence of purely imaginary limits, such as giant-magnon and constant-radii solutions. Therefore, we just have to pay attention to the undeformed limit of~(\ref{Third}) and its degenerate limits with $\zeta_{2}=\zeta_{3}$, $\zeta_{3}=\zeta_{4}$ and $\zeta_{2}=\zeta_{3}=\zeta_{4}$. If (\ref{Two}) is applied to (\ref{Third}), we obtain the solution (\ref{Uno}) with the parameters (\ref{omegalimit}) of the previous subsection. In the degenerate limits with $\zeta_{2}=\zeta_{3}$ and $\zeta_{3}=\zeta_{4}$, we respectively obtain the giant-magnon solution~(\ref{GMIII}) and the constant-radii solution $\zeta=\bar{\zeta}_{3}$. The latter is also obtained when $\zeta_{2}=\zeta_{3}=\zeta_{4}$. Reversing the viewpoint, this result means that the solution~(\ref{Third}), together with its giant-magnon and constant-radii degenerate limits, constitutes the extension of the corresponding undeformed solutions under the presence of purely imaginary deformation parameters.

Consider now the situation where the roots $\zeta_{3}$ and $\zeta_{4}$ are real, and the two remaining roots are complex-conjugated to each other, $\zeta_{2}\,\pm i\,\zeta_{1}$.   
As opposed to the case in which the parameters $\kappa_{\pm}$ are real, there exists just one interval that leads to a bounded solution such that the right-hand side of (\ref{Differential}) is non-negative. 
If we consider the non-degenerate setting with the ordering $\zeta_{3}<\zeta_{4}$, the ellipsoidal coordinate can lie in $\zeta_{3} \le \zeta \le \zeta_{4}$. 
The solution here can be derived upon integration as above if we write the polynomial (\ref{Differential}) as
\be
\label{Refer}
\zeta'^2 = 4\frac{\left(k_{-}\omega_{1}-k_{+}\omega_{2}\right)^2}{\left(\omega_{2}^2-\omega_{1}^2\right)^2}  \left(\zeta_{4}-\zeta\right)\left(\zeta-\zeta_{3}\right)\left[\left(\zeta-\zeta_{2}\right)^2+\zeta_{1}^2\right] \ .
\ee
Therefore, 
\be
\label{Direction}
\begin{split}
&2\frac{\abs{k_{-}\omega_{1}-k_{+}\omega_{2}}}{\omega_{2}^2-\omega_{1}^2}\sigma=\int_{\zeta_{3}}^{\zeta}\frac{\dif\eta}{\sqrt{\left(\zeta_{4}-\eta\right)\left(\eta-\zeta_{3}\right)\left[\left(\eta-\zeta_{2}\right)^2+\zeta_{1}^2\right]}} \\ 
&=\frac{1}{\sqrt{pq}}F\left(2\arccotan_{\left[0,\pi/2\right]}\sqrt{\frac{q\left(\zeta_{4}-\zeta\right)}{p\left(\zeta-\zeta_{3}\right)}},\mu\right) \ ,
\end{split}
\ee 
where we have defined 
\be
\label{Pe}
p = \sqrt{\left(\zeta_{4}-\zeta_{2}\right)^2+\zeta_{1}^2} \ , \quad q = \sqrt{\left(\zeta_{3}-\zeta_{2}\right)^2+\zeta_{1}^2} \ ,
\ee
the elliptic modulus is
\be
\label{Mi}
\mu=\frac{1}{2}\sqrt{\frac{\left(\zeta_{4}-\zeta_{3}\right)^2-\left(p-q\right)^2}{pq}} \ ,
\ee
and we have set to zero the integration constant. The inversion of (\ref{Direction}) leads to
\be
\label{Fourth}
\zeta = \zeta_{3} + \frac{q\left(\zeta_{4}-\zeta_{3}\right)\left(1-\cn\left(2\omega\sigma,\mu\right)\right)^2}{p \sn^2\left(2\omega\sigma,\mu\right)+q\left(1-\cn\left(2\omega\sigma,\mu\right)\right)^2} \ ,
\ee
where we have defined
\be
\label{Omega}
\omega=\frac{\abs{k_{-}\omega_{1}-k_{+}\omega_{2}}}{\omega_{2}^2-\omega_{1}^2}\sqrt{pq} \ .
\ee
To the best of our knowledge, this class of solutions has not appeared in previous works. 
The condition $\omega_{1}^2 \le \zeta \le \omega_{2}^2$ entails that $\omega_{1}^2 \le \zeta_{3} < \zeta_{4} \le \omega_{2}^2$, and $\zeta\left(\sigma\right)=\zeta\left(\sigma+2\pi\right)$ that $\omega$ satisfies a periodicity condition alike (\ref{Relation}) where $\mu$ therein is replaced by the elliptic modulus (\ref{Mi}). In addition, the boundedness condition implies that the constant-radii solutions $\zeta=\zeta_{3}$ and $\zeta=\zeta_{4}$ are also admissible.  

We may cast some light on the new solution by means of the study of its degenerate limits. First of all, we consider the case in which two roots become equal. The most straightforward case is the one in which $\zeta_{3}=\zeta_{4}$, because the range of $\zeta$ shrinks to a point and we are led to $\zeta=\zeta_{3}$. On the contrary, the case involving $\zeta_{1}=0$ is more involved because it gives rise to different limits depending on the hierarchy satisfied by the real roots $\zeta_{3}$ and $\zeta_{4}$ and degenerate real root $\zeta_{2}$. Let us begin with the array $\zeta_{2}<\zeta_{3}<\zeta_{4}$. From (\ref{Pe}) we have
\be
p=\zeta_{4}-\zeta_{2} \ , \quad q=\zeta_{3}-\zeta_{2} \ ,
\ee 
and thus 
\be
\mu=0 \ , \quad \omega = \frac{\abs{k_{-}\omega_{1}-k_{+}\omega_{2}}}{\omega_{2}^2-\omega_{1}^2} \sqrt{ \left(\zeta_{4}-\zeta_{2}\right)\left(\zeta_{3}-\zeta_{2}\right) } \ .
\ee
We note that the periodicity condition still holds with $\omega=n/2$. Accordingly, the Jacobian elliptic functions reduce to their trigonometric counterparts, and the solution becomes
\be
\zeta=\zeta_{3}+\frac{\left(\zeta_{4}-\zeta_{3}\right)\left(\zeta_{3}-\zeta_{2}\right)\sin^2\left(\omega\sigma\right)}{\zeta_{4}-\zeta_{2}-\left(\zeta_{4}-\zeta_{3}\right)\sin^2\left(\omega\sigma\right)} \ .
\ee
If we shift the spatial world-sheet coordinate as $\sigma\mapsto\sigma+\pi/2\omega$, we find the formula (\ref{Limit}). It is worth to underline that performing a shift is equivalent to choosing an alternative non-zero integration constant. The general solution (\ref{Third}) and (\ref{Fourth}) then share the same degenerate limit. This property may traced back to the respective expressions of the polynomials (\ref{noitauqE}) and (\ref{Refer}), which become identical, and thus the solutions bounded between the same pairs of roots should also do so. 
We can bear this result out by the degenerate limit involving the hierarchy $\zeta_{3}<\zeta_{4}<\zeta_{2}$. 
In this case, we find that (\ref{Fourth}) reduces to (\ref{timiL}) once we relabel $\zeta_{2}\mapsto\zeta_{3}$, $\zeta_{4}\mapsto\zeta_{2}$ and $\zeta_{3}\mapsto\zeta_{1}$.

Let us now consider $\zeta_{3}<\zeta_{2}<\zeta_{4}$. In this case, we have
\be
p=\zeta_{4}-\zeta_{2} \ , \quad q=\zeta_{2}-\zeta_{3} \ ,
\ee 
and then
\be
\mu=1 \ , \quad \omega = \frac{\abs{k_{-}\omega_{1}-k_{+}\omega_{2}}}{\omega_{2}^2-\omega_{1}^2} \sqrt{ \left(\zeta_{4}-\zeta_{2}\right)\left(\zeta_{2}-\zeta_{3}\right) } \ .
\ee
Since the elliptic modulus equals one, this limit constitutes an aperiodic limit of the solution (\ref{Fourth}). The periodicity condition of the frequency thus breaks down. If we apply the limit to the solution, it becomes
\be
\zeta=\zeta_{3}+\frac{\left(\zeta_{4}-\zeta_{3}\right)\left(\zeta_{2}-\zeta_{3}\right)\tanh^2\left(\omega\sigma\right)}{\zeta_{4}-\zeta_{2}+\left(\zeta_{2}-\zeta_{3}\right)\tanh^2\left(\omega\sigma\right)} \ .
\ee
We have that the ellipsoidal coordinate ranges between $\zeta_{3}$ and $\zeta_{2}$, but it does not reach $\zeta_{4}$. This expression is the giant-magnon solution (\ref{GMIV}) after the relabelling $\zeta_{3}\mapsto\zeta_{1}$. Besides the violation of the periodicity conditions, the degenerate limit implies that the role of upper bound previously played by $\zeta_{4}$ is transferred to $\zeta_{2}$, which becomes the new nearest real root to $\zeta_{3}$ from above. We can also obtain the  alternative giant-magnon solution (\ref{GMV}) if we proceed along the lines of our previous analysis of aperiodic limits of general solutions. In particular, we may choose a non-vanishing integration constant in~(\ref{Direction}) or, equivalently, shift the spatial world-sheet coordinate before we take the degenerate limit. Following the latter path, we shift the spatial world-sheet coordinate as $\sigma\mapsto\sigma+K\left(\mu\right)/\omega$, which turns (\ref{Fourth}) into
\be
\zeta = \zeta_{4} - \frac{p\left(\zeta_{4}-\zeta_{3}\right)\sn^2\left(2\omega\sigma,\mu\right)}{p\sn^2\left(2\omega\sigma,\mu\right)+q\left(1+\cn\left(2\omega\sigma,\mu\right)\right)^2} \ .
\ee
The degenerate limit now provides
\be
\zeta=\zeta_{4}-\frac{\left(\zeta_{4}-\zeta_{2}\right)\left(\zeta_{4}-\zeta_{3}\right)\tanh^2\left(\omega\sigma\right)}{\zeta_{2}-\zeta_{3}+\left(\zeta_{4}-\zeta_{2}\right)\tanh^2\left(\omega\sigma\right)} \ .
\ee
This equation reproduces (\ref{GMV}) if we relabel $\zeta_{3}\mapsto\zeta_{1}$. The role of lower bound has been transferred from $\zeta_{3}$ to $\zeta_{2}$ since the latter has become the nearest root to $\zeta_{4}$ from below. We can solve this apparently contradictory situation, according to which two different giant-magnon solutions constitute the degenerate limit of the solution (\ref{Fourth}), recalling what we have obtained for the degenerate hierarchy of real roots $\zeta_{1}<\zeta_{2}=\zeta_{3}<\zeta_{4}$. In that case, the coalescence of the intermediate real roots $\zeta_{2}=\zeta_{3}$ provided the giant-magnon solutions (\ref{GMIV}) and (\ref{GMV}) as the aperiodic limit of (\ref{Second}) and (\ref{Third}), respectively. Both solutions were admissible if $\omega_{1}^2\le \zeta_{1} <\zeta_{2} < \zeta_{4} \le \omega_{2}^2$. Taking into account that $\zeta_{3}\mapsto\zeta_{1}$, we puzzled the problem out realizing that the boundedness condition $\omega_{1}^2 \le \zeta_{3} < \zeta_{4} \le \omega_{2}^2$ (assumed to hold in order for the limit of the solution to exist) implies that $\omega_{1}^2 \le \zeta_{2} \le \omega_{2}^2$. The solution involving two real roots and two-complex conjugated roots thus generates both giant-magnon solutions simultaneously. Conversely, we may consider the solution (\ref{Fourth}) as the parent periodic solution of a twofold giant-magnon solution. Moreover, since the degenerate limits of three or four roots in (\ref{Refer}) involve only real roots, they are already comprised in the discussion of the degeneration of the hierarchy of four real roots. 

It remains to study the undeformed limit of the solution to (\ref{Refer}). In the non-degenerate scenario, the limit of vanishing deformation parameters entails the set of roots $\zeta_{2}\,\pm i\,\zeta_{1}$, 
$\zeta_{3}$ and $\zeta_{4}$ should correspond to a set of roots $\bar{\zeta}_{2}\,\pm i\,\bar{\zeta}_{1}$ and $\bar{\zeta}_{3}$ of the cubic polynomial (\ref{Cubic}), with $\bar{\zeta}_{1}$, $\bar{\zeta}_{2}$ and $\bar{\zeta}_{3}$ real.
Such a statement is justified by two facts. First, the impossibility of the non-degenerate set of two real and two complex-conjugated roots to give rise continuously to a non-degenerate set of three real roots. 
Secondly, the fact that the divergence of one real root intervenes in the reduction, since the two non-real roots are intertwined between them by complex-conjugation and hence cannot diverge separately. Regarding the latter remark, the retrieval of the prefactor minus four in (\ref{CubicRoots}) precludes the highest root $\zeta_{4}$ from diverging. Therefore, we should have
\be
\label{refeR}
\zeta_{3} \rightarrow-\infty  \ , \quad  \zeta_{4} \rightarrow \bar{\zeta}_{3} \ , \quad \zeta_{i} \rightarrow \bar{\zeta}_{i} \ , \quad i=1,2, \quad \textnormal{with} \quad
\frac{\left(k_{-}\omega_{1}-k_{+}\omega_{2}\right)^2}{\left(\omega_{2}^2-\omega_{1}^2\right)}\zeta_{3}\rightarrow -1 \ .
\ee
Building on the previous discussions on the undeformed limit, we conclude that (\ref{Fourth}) lacks any undeformed counterpart in the non-degenerate scenario because the limit contravenes the bound $\omega_{1}^2\leq\zeta_{3}$ at some point. While for the same reason the constant-radii solutions $\zeta=\zeta_{3}$ lacks any undeformed limit, the counterpart of solutions $\zeta=\zeta_{4}$ is $\zeta=\bar{\zeta}_{3}$. Furthermore, the degenerate limits of (\ref{Refer}) are connected with the degenerate limits of (\ref{noitauqE}), and hence all the considerations regarding the latter carry over into the former. We merely emphasize that the twofold giant-magnon solution loses the component bounded by the divergent root since it becomes inadmissible when the aforementioned bound is violated.

Let us make a few concluding remarks on the last two scenarios. One of them involves the polynomial with two pairs of complex-conjugated roots. It is enough to observe that this scenario involves a non-positive polynomial in the right-hand side of (\ref{Differential}), 
and thus it does not provide any admissible solution. This is in accordance with the non-degenerate setting in the undeformed limit, because the two pairs of complex-conjugated roots could give neither a set of three real roots, 
nor a set of a real root and a pair of complex-conjugated roots, in which all of them are different. The degenerate limits of this setting are encompassed in previously considered cases.

The other scenario consists in considering that the equality $k_{-}\omega_{1}-k_{+}\omega_{2}=0$ is satisfied. This turns (\ref{noitauqE}) into a cubic polynomial $Q_{3}\left(\zeta\right)$. The coefficient $A$ of the third-order term follows from (\ref{A}) at the end of subsection \ref{Real} after we introduce $\kappa_{\pm}=ik_{\pm}$. Analogously to the case involving real deformation parameters, we find that neither $A$ nor the discriminant of $Q_{3}\left(\zeta\right)$ have definite sign (even for $\left|k_{\pm}\right|<1$). The discussion concluding the subsection \ref{Real} then extends almost identically to the present setting, and hence we omit it.

%%%%%%%%%%%%%%%%%%%%%%%%%%%%%%%%%%%%%%%%%%%%%%%%%%%%%%%%%%%%%%%%%%
%%%%%%%%%%%%%%%%%%%%%%%%%%%%%%%%%%%%%%%%%%%%%%%%%%%%%%%%%%%%%%%%%%

\section{Conclusions}

In this article we have put forward the two-parameter deformation of the Neumann-Rosochatius system for spinning strings that rotate in the three-sphere truncation 
of the double Yang-Baxter deformation of the $\Background$ background. In section~\ref{Dos} we have employed the generalized spinning-string ansatz to obtain 
the Hamiltonian of the underlying mechanical system, which remains integrable. We have also derived an expression for the Uhlenbeck constants assuming 
that the usual constraints for the mechanical system are preserved by the deformation. The argument carries over to any other deformation that preserves the directions 
of the Cartan subalgebra of the three-sphere. In section~\ref{Tres} we have constructed formally the general solution to the system by means of the associated ellipsoidal coordinate, where we have also studied its degenerate and undeformed limits. 
We have differentiated the cases of real and purely imaginary parameters. Admissible solutions depend on the hierarchy of the roots and their reality properties. 
We have shown that aperiodic limits involving the degeneration of two roots renders into giant-magnon and constant-radii solution.
If the deformation parameters are purely imaginary, we have found that periodic trigonometric solutions also emerge when two real roots merge, and we have constructed a new class of solutions when two roots are real and two of them are complex-conjugated to each other.
The latter class reduce to twofold giant-magnon solutions in its degenerate limit of infinite period. 

The most immediate problem suggested by the present text refers to the reduction to a classical mechanical model of spinning strings that rotate in the double Yang-Baxter deformation 
of the three-dimensional anti-de Sitter space. In the undeformed limit it leads to an analytic continuation of the Neumann-Rosochatius system where the oscillators are confined 
to an hyperboloid rather than to a sphere~\cite{NR2}. Since the time coordinate cannot posses winding index, the counterpart of (\ref{Quartic}) would have 
one parameter less than the quartic polynomial considered here. This feature may shed light on the presence of possible limits in which the general solution simplifies, where the attainment of expressions 
for the conserved charges is also more manageable. On the other hand, the legitimate range for the deformation parameters is considerably more involved regarding the anti-de Sitter space~\cite{Hoare}, 
and presumably solutions exhibiting new behavior could emerge. Besides, classical strings pulsating in such a background are also amenable to a treatment in terms of the mechanical system 
at issue. Even though their analysis is utterly analogous to that of spinning strings, the interchange between time and space world-sheet coordinates sharply simplifies the expression for the conserved 
charges~\cite{NR2,Minahan,Engquist}. This fact could facilitate the analysis of the dispersion relation for these solutions.

The Neumann-Rosochatius system does not restrictively concern closed strings. The usage of a rotationally-invariant ansatz in~\cite{DF} allowed minimal surfaces to be represented by means of this mechanical model. 
The classical solutions are again attainable systematically in terms of hyperelliptic functions and limits thereof, but they display their very own characteristics due to the Dirichlet boundary conditions that should be imposed in this case. 
Therefore, a possible line to pursue consists in adapting the approach considered here to study open strings propagating in the double Yang-Baxter deformation of the $\Background$ background. In particular, it should be possible 
to analyze the dominance of the regularized on-shell action of different solutions with the same boundary conditions, and to examine if the deformation parameters could induce a phase transition 
of the Gross-Ooguri type~\cite{Gross}. Certain ambiguities could in fact arise in the computation of the regularized on-shell action, in a parallel fashion to those that appear in the study of the quark-antiquark minimal surface 
in the $\eta$-deformation of the $\Paradigm$ background~\cite{Kameyama}. This would require to extend the criterion on the basis of which divergences are excluded to this scenario. 

\label{Cuatro}

%%%%%%%%%%%%%%%%%%%%%%%%%%%%%%%%%%%%%%%%%%%%%%%%%%%%%%%%%%%%%%%%%%
%%%%%%%%%%%%%%%%%%%%%%%%%%%%%%%%%%%%%%%%%%%%%%%%%%%%%%%%%%%%%%%%%%

\vspace{12mm}

\centerline{\bf Acknowledgments}

\vspace{2mm}

\no
We are grateful to J.~M.~Nieto for correspondence and a careful reading of the manuscript. 
This work is supported by grant PGC2018-095382-B-I00 and by BSCH-UCM through grant GR3/14-A 910770. 
R.~R. acknowledges the support of the Universidad Complutense de Madrid through the predoctoral grant CT42/18-CT43/18. R.~R. also acknowledges 
the organizers of the program \textit{YRISW 2020: A modern primer for superconformal field theories} at DESY  
for support while this work was being completed. 

%%%%%%%%%%%%%%%%%%%%%%%%%%%%%%%%%%%%%%%%%%%%%%%%%%%%%%%%%%%%%%%%%%%%%%%%
%%%%%%%%%%%%%%%%%%%%%%%%%%%%%%%%%%%%%%%%%%%%%%%%%%%%%%%%%%%%%%%%%%%%%%%%

\end{document}